\documentclass[12pt]{article}
\usepackage[top=0.7in,right=0.5in,left=.5in,bottom=0.7in]{geometry}
\usepackage{latexsym}
\usepackage{amsmath,amsfonts,amssymb}
\usepackage{bm}
\usepackage{graphicx}
\usepackage{tabularx}
\usepackage{enumerate}
\usepackage{multirow}
\usepackage{booktabs}
\usepackage[font=footnotesize,labelfont=bf]{caption}
\usepackage{subfig}
\usepackage[svgnames,table]{xcolor}
\usepackage{latexsym}
\usepackage{amsmath}
\usepackage{booktabs}
\usepackage{multirow}
\usepackage{setspace}
\usepackage[toc,page]{appendix}
\usepackage{mathtools}
\usepackage{cancel}
\usepackage{relsize}
\usepackage{empheq}
\usepackage{mathrsfs}
\usepackage{breqn}
\usepackage{arcs}
\usepackage{amssymb}
\usepackage{wasysym}
\usepackage{pifont}
\usepackage{esint}
\usepackage{mathrsfs}

\begin{document}

\newcolumntype{L}[1]{>{\raggedright\let\newline\\\arraybackslash\hspace{0pt}}m{#1}}
\newcolumntype{C}[1]{>{\centering\let\newline\\\arraybackslash\hspace{0pt}}m{#1}}
\newcolumntype{R}[1]{>{\raggedleft\let\newline\\\arraybackslash\hspace{0pt}}m{#1}}

\def\ds{\displaystyle}

\newcommand{\beq}{\begin{equation}}
\newcommand{\eeq}{\end{equation}}
\newcommand{\lb}{\label}
\newcommand{\beqar}{\begin{eqnarray}}
\newcommand{\eeqar}{\end{eqnarray}}
\newcommand{\barr}{\begin{array}}
\newcommand{\earr}{\end{array}}
\newcommand{\jump}{\parallel}
 \newcommand{\varbeta}

\def\c{{\circ}}

\newcommand{\Ehat}{\hat{E}}
\newcommand{\That}{\hat{\bf T}}
\newcommand{\Ahat}{\hat{A}}
\newcommand{\chat}{\hat{c}}
\newcommand{\shat}{\hat{s}}
\newcommand{\khat}{\hat{k}}
\newcommand{\muhat}{\hat{\mu}}
\newcommand{\mc}{M^{\scriptscriptstyle C}}
\newcommand{\mei}{M^{\scriptscriptstyle M,EI}}
\newcommand{\mec}{M^{\scriptscriptstyle M,EC}}
\newcommand{\hbeta}{{\hat{\beta}}}
\newcommand{\rec}[2]{\left( #1 #2 \ds{\frac{1}{#1}}\right)}
\newcommand{\rep}[2]{\left( {#1}^2 #2 \ds{\frac{1}{{#1}^2}}\right)}
\newcommand{\derp}[2]{\ds{\frac {\partial #1}{\partial #2}}}
\newcommand{\derpn}[3]{\ds{\frac {\partial^{#3}#1}{\partial #2^{#3}}}}
\newcommand{\dert}[2]{\ds{\frac {d #1}{d #2}}}
\newcommand{\dertn}[3]{\ds{\frac {d^{#3} #1}{d #2^{#3}}}}
\newcommand{\btimes}{\mathbin{\rotatebox[origin=c]{90}{$\ltimes$}}}

\def\bob{{\, \underline{\overline{\otimes}} \,}}
\def\ob{{\, \underline{\otimes} \,}}
\def\scalp{\mbox{\boldmath$\, \cdot \, $}}
\def\gdp{\makebox{\raisebox{-.215ex}{$\Box$}\hspace{-.778em}$\times$}}
\def\daa{\makebox{\raisebox{-.050ex}{$-$}\hspace{-.550em}$: ~$}}
\def\mK{\mbox{${\mathcal{K}}$}}
\def\cK{\mbox{${\mathbb {K}}$}}
\def\ellipse{\raisebox{-.5pt}{\scalebox{.5}[1.5]{$\circ$}}}

\DeclarePairedDelimiter{\abso}{\lvert}{\rvert}
\DeclarePairedDelimiter{\norma}{\lVert}{\rVert}

\def\Xint#1{\mathchoice
   {\XXint\displaystyle\textstyle{#1}}%
   {\XXint\textstyle\scriptstyle{#1}}%
   {\XXint\scriptstyle\scriptscriptstyle{#1}}%
   {\XXint\scriptscriptstyle\scriptscriptstyle{#1}}%
   \!\int}
\def\XXint#1#2#3{{\setbox0=\hbox{$#1{#2#3}{\int}$}
     \vcenter{\hbox{$#2#3$}}\kern-.5\wd0}}
\def\ddashint{\Xint=}
\def\fpint{\Xint=}
\def\dashint{\Xint-}
\def\cpvint{\Xint-}
\def\intl{\int\limits}
\def\cpvintl{\cpvint\limits}
\def\fpintl{\fpint\limits}
\def\ointl{\oint\limits}
\def\bA{{\bf A}}
\def\ba{{\bf a}}
\def\bB{{\bf B}}
\def\bb{{\bf b}}
\def\bc{{\bf c}}
\def\bC{{\bf C}}
\def\bD{{\bf D}}
\def\bE{{\bf E}}
\def\be{{\bf e}}
\def\bbf{{\bf f}}
\def\bF{{\bf F}}
\def\bG{{\bf G}}
\def\bg{{\bf g}}
\def\bi{{\bf i}}
\def\bH{{\bf H}}
\def\bK{{\bf K}}
\def\bL{{\bf L}}
\def\bM{{\bf M}}
\def\bN{{\bf N}}
\def\bn{{\bf n}}
\def\b0{{\bf 0}}
\def\bo{{\bf o}}
\def\bX{{\bf X}}
\def\bx{{\bf x}}
\def\bP{{\bf P}}
\def\bp{{\bf p}}
\def\bQ{{\bf Q}}
\def\bq{{\bf q}}
\def\bR{{\bf R}}
\def\bS{{\bf S}}
\def\bs{{\bf s}}
\def\bT{{\bf T}}
\def\bt{{\bf t}}
\def\bU{{\bf U}}
\def\bu{{\bf u}}
\def\bv{{\bf v}}
\def\bw{{\bf w}}
\def\bW{{\bf W}}
\def\by{{\bf y}}
\def\bz{{\bf z}}
\def\T{{\bf T}}
\def\Te{\textrm{T}}
\def\Id{{\bf I}}
\def\bxi{\mbox{\boldmath${\xi}$}}
\def\balpha{\mbox{\boldmath${\alpha}$}}
\def\bbeta{\mbox{\boldmath${\beta}$}}
\def\bepsilon{\mbox{\boldmath${\epsilon}$}}
\def\bvarepsilon{\mbox{\boldmath${\varepsilon}$}}
\def\bomega{\mbox{\boldmath${\omega}$}}
\def\bphi{\mbox{\boldmath${\phi}$}}
\def\bsigma{\mbox{\boldmath${\sigma}$}}
\def\bfeta{\mbox{\boldmath${\eta}$}}
\def\bDelta{\mbox{\boldmath${\Delta}$}}
\def\btau{\mbox{\boldmath $\tau$}}
\def\tr{{\rm tr}}
\def\dev{{\rm dev}}
\def\div{{\rm div}}
\def\Div{{\rm Div}}
\def\Grad{{\rm Grad}}
\def\grad{{\rm grad}}
\def\Lin{{\rm Lin}}
\def\Sym{{\rm Sym}}
\def\Skw{{\rm Skew}}
\def\abs{{\rm abs}}
\def\Re{{\rm Re}}
\def\Im{{\rm Im}}
\def\capB{\mbox{\boldmath${\mathsf B}$}}
\def\capC{\mbox{\boldmath${\mathsf C}$}}
\def\capD{\mbox{\boldmath${\mathsf D}$}}
\def\capE{\mbox{\boldmath${\mathsf E}$}}
\def\capG{\mbox{\boldmath${\mathsf G}$}}
\def\tcapG{\tilde{\capG}}
\def\capH{\mbox{\boldmath${\mathsf H}$}}
\def\capK{\mbox{\boldmath${\mathsf K}$}}
\def\capL{\mbox{\boldmath${\mathsf L}$}}
\def\capM{\mbox{\boldmath${\mathsf M}$}}
\def\capR{\mbox{\boldmath${\mathsf R}$}}
\def\capW{\mbox{\boldmath${\mathsf W}$}}
\def\dellipse{\ooalign{\raise-.25pt\hbox{\ellipse}\cr\ellipse}}

\def\i{\mbox{${\mathrm i}$}}
\def\mC{\mbox{\boldmath${\mathcal C}$}}
\def\mB{\mbox{${\mathcal B}$}}
\def\mE{\mbox{${\mathcal{E}}$}}
\def\mL{\mbox{${\mathcal{L}}$}}
\def\mK{\mbox{${\mathcal{K}}$}}
\def\mV{\mbox{${\mathcal{V}}$}}
\def\C{\mbox{\boldmath${\mathcal C}$}}
\def\E{\mbox{\boldmath${\mathcal E}$}}

\def\ACME{{ Arch. Comput. Meth. Engng.\ }}
\def\ARMA{{ Arch. Rat. Mech. Analysis\ }}
\def\AMR{{ Appl. Mech. Rev.\ }}
\def\ASCEEM{{ ASCE J. Eng. Mech.\ }}
\def\acta{{ Acta Mater. \ }}
\def\CMAME {{ Comput. Meth. Appl. Mech. Engrg.\ }}
\def\CRAS{{ C. R. Acad. Sci., Paris\ }}
\def\EFM{{ Eng. Fract. Mech.\ }}
\def\EJMA{{ Eur.~J.~Mechanics-A/Solids\ }}
\def\IJES{{ Int. J. Eng. Sci.\ }}
\def\IJF{{ Int. J. Fracture\ }}
\def\IJMS{{ Int. J. Mech. Sci.\ }}
\def\IJNAMG{{ Int. J. Numer. Anal. Meth. Geomech.\ }}
\def\IJP{{ Int. J. Plasticity\ }}
\def\IJSS{{ Int. J. Solids Structures\ }}
\def\IngA{{ Ing. Archiv\ }}
\def\JAM{{ J. Appl. Mech.\ }}
\def\JAP{{ J. Appl. Phys.\ }}
\def\JE{{ J. Elasticity\ }}
\def\JM{{ J. de M\'ecanique\ }}
\def\JMPS{{ J. Mech. Phys. Solids\ }}
\def\JoMMS{{ J. Mech. Materials Structures\ }}
\def\Macro{{ Macromolecules\ }}
\def\MOM{{ Mech. Materials\ }}
\def\MMS{{ Math. Mech. Solids\ }}
\def\MMT{{ Metall. Mater. Trans. A}}
\def\MPCPS{{ Math. Proc. Camb. Phil. Soc.\ }}
\def\MRC{{ Mech. Res. Comm.}}
\def\MSE{{ Mater. Sci. Eng.}}
\def\PMPS{{ Proc. Math. Phys. Soc.\ }}
\def\PRE{{ Phys. Rev. E\ }}
\def\PRL{{ Phys. Rev. Letters\ }}
\def\PRSL{{ Proc. R. Soc.\ }}
\def\rock{{ Rock Mech. and Rock Eng.\ }}
\def\QAM{{ Quart. Appl. Math.\ }}
\def\QJMAM{{ Quart. J. Mech. Appl. Math.\ }}
\def\SCRMAT{{ Scripta Mater.\ }}
\def\SM{{\it Scripta Metall. }}

\def\salto#1#2{
[\mbox{\hspace{-#1em}}[#2]\mbox{\hspace{-#1em}}]}

\title{Torsion of elastic solids with sparse voids parallel to the twist axis}\date{}

\author{S. Shahzad$^1$ and F. Dal Corso$^2$\\
$^1$Department of Civil Engineering, Aalto University, Helsinki, Finland\\
$^2$DICAM, University of Trento, via Mesiano 77, I-38123 Trento, Italy}

\maketitle

\begin{abstract}
\noindent

With the purpose of investigating a linear elastic solid containing
a dilute distribution of cylindrical and prismatic holes parallel to the torsion axis,
the  full-field solution for an infinite elastic plane containing a single void and subject to torsion  is derived.
The obtained  solution is exploited to derive the analytic expressions for the Stress Concentration Factor (SCF) related to the presence of an
elliptical hole, for the Stress Intensity Factor (SIF) for
hypocycloidal-shaped hole and star-shaped cracks, and for  the Notch Stress
Intensity Factor (NSIF) for star-shaped polygons. Special sets of the void location are obtained for which peculiar mechanical behaviours are displayed, such as the stress annihilation at some points along the  boundary of elliptical voids and the stress
singularity removal at the cusps/points of hypocycloidal shaped/isotoxal star-shaped polygonal voids. By means of finite element simulations it is finally shown that
the presented closed-form expressions for the stress intensification  provide reliable predictions even for finite domain realizations and, in particular,
the infinite-plane solution remains highly accurate when the size of smooth and non-smooth
external boundary is greater than twice and five times the void dimension, respectively. Under these geometrical conditions, the derived analytical expressions represent a valid \lq  guide tool' in mechanical design.
\end{abstract}


\section{Introduction}

The strong intensification of the stress fields around inhomogeneities, flaws, defects, cracks and voids inside a solid represents a fundamental aspect in mechanical design, being strictly connected to the material strength and its failure.
For this reason, stress intensification has been the subject of an intensive research activity, encompassing analytical \cite{amenyah, craciun, roaz,  savin, willis},
numerical \cite{barbieri,num2,num2b,num4,num1,num3,num5}, and experimental \cite{gross,exp2,inclusioni,noselli,ostervig,rosakis,exp1}  approaches.

Beside the strong research effort in planar and three-dimensional problems of elasticity,
relatively little attention has been devoted to the stress intensification in  inhomogeneous solids subject to torsional loading.
Within this framework,  the research has been mainly focussed on (i.) cylindrical shafts containing notches of different geometries \cite{suzuki, zappa0,zappa1,zappa2}; (ii.) cylinders with elliptical cross-section containing a single crack \cite{iran2,sih};
(iii.) circular shafts with quasi-regular polygonal voids  \cite{iran}; (iv.) 
cross sections weakened by edge cracks \cite{iran2,kuli,sih,tweed}; (v.) neutrality of coated cavities of various shapes in cylinders with elliptical cross-section \cite{wangwang}; (vi.) multiply connected domains \cite{bartels,chen0,chen1,poland,iran2,jan,li, lin}; and (vii.) star-shaped cracks in square and circular cross-sections \cite{chenyz1,chenyz2}. However, except for some special geometry, the results are
usually obtained through the implementation of numerical techniques and  very few analytical expressions are currently available as a \lq  guide tool' for engineers.

The present investigation aims to provide a new class of analytical expressions for torsion problems. In particular, with reference to a linear elastic solid containing
a dilute distribution of cylindrical and prismatic holes parallel to the torsion axis, the  full-field solution for an infinite elastic plane containing a single void and subject to torsion  is obtained in a closed-form
expression by means of complex potential technique and conformal mapping \cite{sokolnikoff}. The solution is obtained for three specific void geometries: ellipse, $n$-cusped hypocycloid, and
$n$-pointed isotoxal star-shaped polygon, the latter including the special cases of
$n$-pointed regular polygon, of $n$-pointed regular star polygon, and of $n$- pointed star-shaped crack.
The achieved solution is exploited to derive  analytical expressions for the Stress Concentration Factor (SCF), Stress Intensity Factor (SIF), and Notch Stress Intensity Factor (NSIF), useful in the evaluation of the stress intensification
along the boundary of an elliptical void, at the cusps of an hypocycloidal-shaped void, at the tips of a star-shaped crack, and at the points of a star-shaped polygonal hole. Similarly to recent results obtained within a \lq pure' out-of-plane setting \cite{partI, partII, cusp, annuli},
 special sets of the void location with respect to the torsion axis have been identified  for which peculiar mechanical behaviours are displayed, such as the stress annihilation at some points along the  boundary of elliptical voids and the stress
singularity removal at the cusps/points of hypocycloidal shaped/isotoxal star-shaped polygonal voids.

Finally, with the purpose of facilitating the application of the presented results to practical realizations, finite element simulations (in Comsol Multiphysics$^\copyright$)  have been performed in order
to assess the influence of the shape and the size of an enclosing elastic finite, not infinite, domain and the related variation in the stress intensification measures from the analytical predictions, obtained under the assumption of infinite elastic domain.
The Finite Element simulations (and results available in literature for some special case) show that the analytic expressions for the stress intensification derived for an infinite domain provide reliable predictions even for finite domains and,
in particular, a great accuracy is achieved when the size of smooth and non-smooth external boundaries is greater than twice and
five times the void dimension, respectively.

\section{Problem formulation and governing equations} \lb{gogo}

\begin{figure}[h!]
  \begin{center}
\includegraphics[width=14 cm]{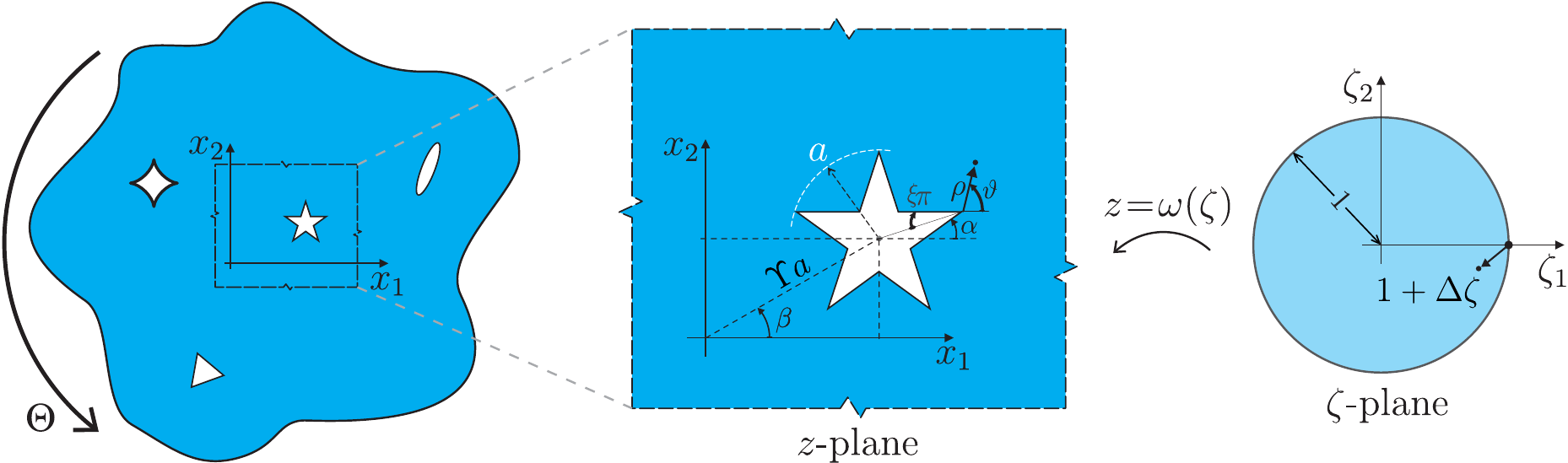}
\caption{(Left) Cross section of a solid subject to a twist angle (per unit length) $\Theta$ and containing a dilute distribution of cylindrical
or prismoidal cavities parallel to the torsion axis $x_3$, orthogonal to the plane realized by the axes $x_1$ and $x_2$. (Center) Due to the considered geometrical assumptions,
the mechanical problem can be investigated as the twist of an infinite elastic plane containing a single void.
The location of the void, enclosed by the smallest circle with radius $a$, can be described through the parameter $\Upsilon$ defining the radial distance $\Upsilon a$ of the void centroid
from the origin of the $x_1-x_2$ reference system and the angles $\beta$ and $\alpha$. (Right)  The transformed complex variable $\zeta$ identifying
the point position within the unit disk in the conformal plane is mapped onto the complex variable $z$ within the physical plane through the conformal mapping function $\omega(\zeta)=z$.
\small
}
\label{fig1}
 \end{center}
\end{figure}

A dilute distribution of voids, with shape of prisms  or of cylinders, is considered within a linear elastic isotropic solid subject to torsion. The voids are considered parallel to each other
and to the torsion axis, corresponding to the axis $x_3$, so that the cross section (realized by the plane $x_1-x_2$, orthogonal to $x_3$) is uniform, Fig. \ref{fig1} (left), namely, independent of the coordinate $x_3$.
The  voids are considered diluted, so that the possible interactions between the voids and with the external boundary of the bar
are disregarded and the elastic problem can be modeled as the twist of a bar with an infinite
cross section (the plane $x_1-x_2$) containing a single void, Fig. \ref{fig1} (center).

For twist loading condition, the displacement components $v_j$  (along the direction $x_j$, $j=1,2,3$) are given by the following expressions  \cite{sokolnikoff}
    \beq
        \label{eq_displacement_field_torsion}
            v_1(x_2,x_3)=-\Theta x_{2} x_{3}\,, \qquad
            v_2(x_1,x_3)=\Theta x_{1} x_{3}\,,  \qquad
            v_3(x_1,x_2)=\Theta \varphi(x_{1},x_{2})\,,
    \eeq
where $\Theta$ is the angle of twist per unit length while $\varphi(x_{1},x_{2})$ is the warping function, to be evaluated, and depending on the cross section geometry.

With reference to the displacement field (\ref{eq_displacement_field_torsion}) and considering a isotropic linear elastic behaviour,
the only non-null strain and stress components are $\gamma_{j3}$ and $\tau_{j3}$ with $j=1,2$. These quantities are defined by
the following kinematical and constitutive relations
  \beq
        \label{eq_shear_strains}
            \gamma_{j3}=\frac{\partial v_3}{\partial x_{j}}+\frac{\partial v_j}{\partial x_{3}}\,,    \qquad
            \tau_{j3}=\mu \gamma_{j3}\,,    \qquad j=1,2,
  \eeq
with $\mu$ being the shear modulus, and considering  the displacement field~\eqref{eq_displacement_field_torsion},
the shear stress components can be obtained  as
    \beq
        \label{eq_stress_field_torsion}
            \tau_{13}=\mu\Theta\left(\frac{\partial\varphi}{\partial x_{1}}-x_{2}\right)\,                              ,\qquad
            \tau_{23}=\mu\Theta\left(\frac{\partial\varphi}{\partial x_{2}}+x_{1}\right)\,.
    \eeq

Therefore under these torsion loading conditions, similarly to Mode III, one of the three eigenvalues of the stress tensor is zero while the
other two have the same absolute value given by
\beq
\label{eq_shear_stress}
\tau=\sqrt{\tau_{13}^2+\tau_{23}^2}.
\eeq

Restricting the attention to quasi-static conditions and negligible body forces, the equilibrium is achieved when the shear stress components satisfy the following differential equation
  \beq
        \label{eq_equilibrium}
            \frac{\partial \tau_{13}}{\partial x_{1}} + \frac{\partial \tau_{23}}{\partial x_{2}}=0\, ,
    \eeq
implying that the warping function $\varphi(x_{1},x_{2})$ is harmonic, $\nabla^2\varphi=0$. Due to the harmonicity of the warping function,
an analytic function $G(z)$ of the complex variable $z=x_{1}+\i x_{2}$ (where $\i$ is the imaginary unit) can be introduced as follows
    \beq
        \label{eq_analytic_function_torsion}
            G(z)=\varphi(x_{1},x_{2})+\i \psi(x_{1},x_{2})\,,
    \eeq
where $\psi(x_{1},x_{2})$ is the conjugate harmonic function (also known as stress function) of the warping function $\varphi(x_{1},x_{2})$, connected each other through the Cauchy-Riemann equations
    \beq
        \label{eq_cauchy_riemann_torsion}
            \frac{\partial \varphi}{\partial x_{1}}=\frac{\partial \psi}{\partial x_{2}}\,                              ,   \qquad
            \frac{\partial \psi}{\partial x_{1}}=-\frac{\partial \varphi}{\partial x_{2}}\, .
    \eeq
Considering the Cauchy-Riemann equations (\ref{eq_cauchy_riemann_torsion}),  the stress function $\psi(x_{1},x_{2})$  can be related to the shear stress components (\ref{eq_stress_field_torsion}) through
\beq
        \label{eq_stress_field_torsion_2}
            \tau_{13}=\mu\Theta\left(\frac{\partial\psi}{\partial x_{2}}-x_{2}\right)\,                             ,\qquad
            \tau_{23}=\mu\Theta\left(-\frac{\partial\psi}{\partial x_{1}}+x_{1}\right)\, ,
    \eeq
and is governed by
\beq\label{nabla_psi}
\nabla^{2}\psi =0.
\eeq

The stress-free boundary condition holding along the void surface $\mathcal{S}$ is defined by $\tau_{j3} n_j=0$ ($j=1,2$), where the summation convention is assumed and $n_j$ is
the component along the $x_j$-axis of the outward normal to the void boundary $\mathcal{S}$. With reference to the stress function $\psi(x_{1},x_{2})$,
the stress-free condition can be read as
\beq
\psi=\dfrac{1}{2}\left(x_1^2+x_2^2\right)  \qquad \text{on} \quad \mathcal{S},
\eeq
which represents together with eqn (\ref{nabla_psi}) the celebrated Dirichlet boundary value problem for the stress function $\psi(x_{1},x_{2})$.

Following the technique introduced by Sokolnikoff \cite{sokolnikoff}, the torsion problem of an infinite plane containing a void inclusion is
 treated by means of conformal mapping. In this technique,
the points of the infinite plane (identified with the complex variable $z=x_1+\i x_2$) are mapped  onto the region inside a unit disk in the
conformal plane (where the position is given by the transformed complex variable $\zeta$, with $\zeta\leq 1$, Fig. \ref{fig1}, right) through the relation
    \beq
        z=\omega(\zeta).
    \eeq
The complex potential $G(z)=G(x_1,x_2)$ can be described within the conformal plane as
\beq
g(\zeta)=G(z)=G(\omega(\zeta)),
\eeq
to be achieved by imposing the stress-free boundary condition on the unit-disk boundary. After some mathematical manipulation, this condition leads to \cite{sokolnikoff}
    \beq
        \label{eq_Schwarz_poisson_formula_final}
            g(\zeta)=\frac{1}{2\pi}\oint_{|\sigma|=1}\frac{\omega(\sigma)\overline{\omega(           \sigma)}}{\sigma-\zeta}d\sigma\,,
    \eeq
    to be solved applying the following property holding for points $z$ inside the unit disk and provided by Cauchy's integral formulae \cite{sokolnikoff},
    \beq
    \label{cauchycauchy}
    \frac{1}{2\pi \i}\oint_{|\sigma|=1}\frac{\sigma^p}{\sigma-\zeta}d\sigma\,=
    \left\{
    \begin{array}{llll}
     0\qquad & \mbox{for}\qquad p <0,\\[2mm]
     1 \qquad & \mbox{for}\qquad p =0,\\[2mm]
     \zeta^p \qquad & \mbox{for}\qquad p> 0.
    \end{array}
    \right.
    \eeq

Once the complex potential $g(\zeta)$ is computed through eqn (\ref{eq_Schwarz_poisson_formula_final}), the shear stress and the out-of-plane displacement can be finally evaluated as
    \beq
        \label{eq_Stress_field_torsion_final}
            \tau_{13}-\i \tau_{23}=\mu\Theta\left[\frac{g^\prime(\zeta)}{\omega^\prime(\zeta)}-\i\overline{\omega(\zeta)}\right]\,, \quad v_3=\Theta \, \Re\left[g(\zeta)\right],
    \eeq
with $\Re[\cdot]$ being the real part of the relevant argument.

\begin{figure}[h!]
  \begin{center}
\includegraphics[width=16 cm]{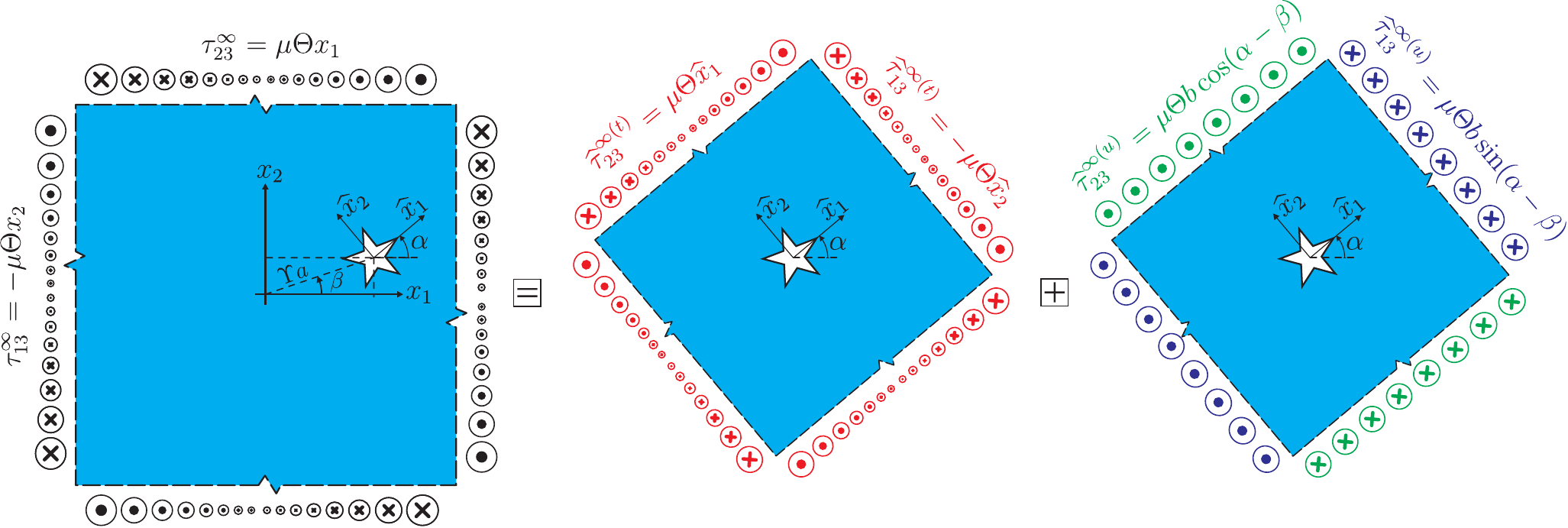}
\caption{By means of  the superposition principle, the generic torsion problem (left) can be decomposed with reference to the $\widehat{x}_{1}-\widehat{x}_{2}$ reference system
as the sum of a remote \lq pure' torsion (having a null unperturbed stress state at the centroid of the void, center)
and a remote uniform Mode III (right). 
\small
}
\label{fig2}
 \end{center}
\end{figure}

In the following analysis, the introduction of a further reference system $\widehat{x}_1-\widehat{x}_2$ is instrumental. The origin of this
system is defined by the centroid of the void, with coordinates $x_1=\Upsilon a\cos\beta$ and $x_2=\Upsilon a \sin\beta$, being  $a$ the radius of the smallest circle  enclosing the void. The system $\widehat{x}_1-\widehat{x}_2$ is parallel and orthogonal to principal axes of inertia of the void,
so that is inclined at a counter-clockwise angle $\alpha$ with respect to the reference system $x_1-x_2$, Fig \ref{fig2} (left), so that the two systems are connected through the following linear
relation
\beq
\label{xstarnostar}
\left[
\begin{array}{lll}
 x_{1}\\
  x_{2}
\end{array}
\right]
=\ds  \left[
\begin{array}{lll}
 \cos\beta\\
  \sin\beta
\end{array}
\right] \Upsilon a +
\left[
   \begin{array}{ccc}
   \cos\alpha\qquad &-\sin\alpha\\
   \sin\alpha\qquad &\cos\alpha
   \end{array}
   \right] \left[
\begin{array}{lll}
 \widehat x_{1}\\
  \widehat x_{2}
\end{array}
\right].
\eeq
The shear stress components $\widehat{\tau}_{13}$
and $\widehat{\tau}_{23}$ referred within the reference system $\widehat{x}_1-\widehat{x}_2$ can be obtained from the shear stress components $\tau_{13}$ and $\tau_{23}$
(referred within the reference system $x_1-x_2$) through the following rotation relationship

\beq
 \widehat\tau_{1 3}-\i  \widehat\tau_{2 3} = \left(\tau_{1 3}-\i \tau_{2 3}\right) e^{\i \alpha}.
\eeq

\subsection{Stress singularities}\label{stresssing}
When the cross section of the void is a polygon or an hypocycloid, shear stress singularity may arise under remote twist
at the tips/cusps of the void, of which asymptotic behaviour can be derived.
The asymptotic fields for this problem can be expressed similarly to those of Mode III \cite{partI}, because the loading conditions  differ in the two cases only for
the in-plane displacements $u_1$ and $u_2$. More in particular, while the in-plane displacements are null in Mode III, these
are non-null in the torsion problem but non-singular, eqn (\ref{eq_displacement_field_torsion}).
Therefore, the shear stress fields $\widehat{\tau}_{j3}$ $(j=1,2)$ expressed within the cartesian
system $\widehat{x}_1-\widehat{x}_2$, with the axis  $\widehat{x}_1$ being
the bisector line of the inclusion vertex (Fig. \ref{fig2}), may be represented by the leading-order term  \cite{andrzei}

\beq
\label{asym_poly_z_star_generale_1}
\left[
\begin{array}{lll}
 \widehat\tau_{1 3}(\rho, \vartheta)\\
   \widehat\tau_{2 3}(\rho, \vartheta)
\end{array}
\right]
\simeq \ds \rho^\lambda
\left[
   \begin{array}{lll}
   \sin\lambda\left(\vartheta-\alpha\right)\\
   \cos\lambda\left(\vartheta-\alpha\right)
   \end{array}
   \right],
\eeq

where $\rho$ is the radial distance from the considered inclusion vertex (Fig. \ref{fig1}, center), $\vartheta$ measures the counter-clockwise angle from the axis $x_1$
(with $\vartheta=\alpha$ for the points within the elastic material along the bisector line of the inclusion vertex,
$\widehat{x}_1$), and $\lambda$ is a parameter defining the stress singularity order, given by
\beq
\lambda=-\frac{1-2\xi}{2(1-\xi)},
\eeq
where $\xi$ defines the angle $2\xi\pi$ interior to the sharp vertex of the void.

The measure of the stress intensity in the case of singular stress fields (useful to detect possible failure conditions)
is provided by the Stress Intensity Factors (SIFs) for the case of cracks ($\xi=0$) and the Notch Stress Intensity Factors (NSIFs) for the case of isotoxal star-shaped
polygonal voids ($\xi\neq 0$), which can be defined as \cite{partI}
\beq
\label{eq_def_nsif}
K_{\textup{III}}=\lim_{\rho
\rightarrow 0} \sqrt{2\pi} \,\rho^{-\lambda}\,\, \widehat{\tau}_{23}(\rho,\vartheta=\alpha).
\eeq

\subsection{Loading decomposition}\label{loaddec}
It is worth remarking that,  due to linearity of the governing equation (\ref{eq_equilibrium}),
the mechanical fields  (\ref{eq_displacement_field_torsion}) and  (\ref{eq_stress_field_torsion})  can be considered as the result of the  superposition of
two \lq simple' remote loadings referred to the system $\widehat{x}_1-\widehat{x}_2$  (Fig. \ref{fig2}):
\begin{itemize}
\item a remote torque with torsion axis coincident with the centroidal axis of the void (Fig. \ref{fig2}, center), namely the origin of the system $\widehat{x}_1-\widehat{x}_2$.
Solution of this problem is referred through the apex ($t$);
\item a remote uniform\footnote{It is worth remarking that, differently from classical problems in fracture mechanics where
a remote linear out-of-plane displacement field is imposed, $\widehat{v}^\infty_3(\widehat{x}_1,\widehat{x}_2)=\widehat{\gamma}^\infty_{13} \widehat{x}_1+\widehat{\gamma}^\infty_{23} \widehat{x}_2 $,
the uniform loading mode III considered in
this investigation (as part of the decomposition of the generic loading)
is realized through remote linear in-plane components for the displacement field, $\widehat{v}^\infty_1(\widehat{x}_3)=\widehat{\gamma}^\infty_{13} \widehat{x}_3$
and $\widehat{v}^\infty_2(\widehat{x}_3)=\widehat{\gamma}^\infty_{23} \widehat{x}_3$. Indeed, differing only for a rigid-body motion,
the two remote conditions are equivalent and corresponding to the same uniform remote stress,
defined by the only non-null components $\widehat{\tau}^\infty_{13}=\mu\widehat{\gamma}^\infty_{13}$ and $\widehat{\tau}^\infty_{23}=\mu\widehat{\gamma}^\infty_{23}$.}
 Mode III (Fig. \ref{fig2}, right), having a non-null magnitude
 whenever the void centroid does not correspond with the origin of the $x_1-x_2$ reference system, $\Upsilon\neq 0$.
Solution of this problem is referred through the apex ($u$).
\end{itemize}
Considering such a decomposition, the displacement field (\ref{eq_displacement_field_torsion}) can be expressed as a function of the position $\widehat{x}_1$ and
$\widehat{x}_2$ and in this reference system as
    \beq
        \label{eq_displacement_field_torsion2}
            \widehat{v}_k=\widehat{v}^{(t)}_k+\widehat{v}^{(u)}_k\,,\,\qquad k=1,2,3,
    \eeq
where
            \beq
        \label{eq_displacement_field_torsion3}
        \begin{array}{lll}
            \widehat{v}_1^{(t)}(\widehat{x}_2,x_3)=-\Theta \widehat{x}_{2} x_{3}\,, \\
            \widehat{v}_2^{(t)}(\widehat{x}_1,x_3)=\Theta \widehat{x}_{1} x_{3}\,,  \\
            \widehat{v}_3^{(t)}(\widehat{x}_1,\widehat{x}_2)=\Theta \widehat{\varphi}^{(t)}(\widehat{x}_{1},\widehat{x}_{2})\,,
            \end{array} \qquad\mbox{and}\qquad
                    \begin{array}{lll}
            \widehat{v}_1^{(u)}(\widehat{x}_2,x_3)=\Theta \Upsilon a \sin (\alpha-\beta)x_3, \\
            \widehat{v}_2^{(u)}(\widehat{x}_1,x_3)=\Theta \Upsilon a \cos (\alpha-\beta)x_3,  \\
            \widehat{v}_3^{(u)}(\widehat{x}_1,\widehat{x}_2)=\Theta \widehat{\varphi}^{(u)}(\widehat{x}_{1},\widehat{x}_{2})\, .
            \end{array}
    \eeq
Consequently, the shear stresses are decomposed as
\beq
        \label{eq_displacement_field_torsion2b}
\widehat{\tau}_{j3}=\widehat{\tau}^{(t)\infty}_{j3}+\widehat{\tau}^{(t)p}_{j3}+\widehat{\tau}^{(u)\infty}_{j3}+\widehat{\tau}^{(u)p}_{j3} \qquad j=1,2,
    \eeq
where the apexes $\infty$ and $p$ define the unperturbed and perturbed fields, respectively, which are given by
        \beq
        \label{eq_displacement_field_torsion4}
        \begin{array}{lll}
            &\widehat{\tau}_{13}^{(t)\infty}(\widehat{x}_2,x_3)=-\mu\Theta\widehat{x}_2,\qquad\qquad
            &\widehat{\tau}_{13}^{(u)\infty}(\widehat{x}_2,x_3)=\mu\Theta\ds \Upsilon a \sin (\alpha-\beta)\,,
              \\[4mm]
            &\widehat{\tau}_{23}^{(t)\infty}(\widehat{x}_1,x_3)=\mu\Theta\widehat{x}_1,
            \qquad\qquad
            &\widehat{\tau}_{23}^{(u)\infty}(\widehat{x}_1,x_3)=\mu\Theta\ds \Upsilon a \cos (\alpha-\beta),
            \end{array}
    \eeq
    showing that the \textit{unperturbed uniform} stress field is  only dependent on the radial distance $\Upsilon$ and on the angular difference $(\alpha-\beta)$, and by
        \beq
        \label{eq_displacement_field_torsion4b}
        \begin{array}{llll}
            &\widehat{\tau}_{13}^{(t)p}(\widehat{x}_2,x_3)=\mu\Theta\ds\frac{\partial \widehat{\varphi}^{(t)}}{\partial \widehat{x}_1},
            \qquad\qquad
            &\widehat{\tau}_{13}^{(u)p}(\widehat{x}_2,x_3)=\mu\Theta\ds\frac{\partial \widehat{\varphi}^{(u)}}{\partial \widehat{x}_1}\,,
              \\[4mm]
            &\widehat{\tau}_{23}^{(t)p}(\widehat{x}_1,x_3)=\mu\Theta\ds\frac{\partial \widehat{\varphi}^{(t)}}{\partial \widehat{x}_2}\,,
            \qquad\qquad
            &\widehat{\tau}_{23}^{(u)p}(\widehat{x}_1,x_3)=\mu\Theta\ds\frac{\partial \widehat{\varphi}^{(u)}}{\partial \widehat{x}_2}.
            \end{array}
    \eeq
The introduced decomposition  provides a further insight in the understanding of the full-field solution and the SIFs and NSIFs values
at varying the distance $\Upsilon$ and the inclination  $(\alpha-\beta)$ of the inclusion with
respect to the $x_1-x_2$ reference system. More in particular, from the above presented decomposition for
mechanical fields, it follows that the intensity factors can be considered as the sum of the intensification
due to the \lq pure' torsion loading condition ($t$) and that due to the uniform Mode III  ($u$),
\beq
K_{\textup{III}}=K^{(t)}_{\textup{III}}+K^{(u)}_{\textup{III}}.
\eeq
Therefore, an additional result of the present analysis is the independent  confirmation of the expression
recently obtained  intensity factor $K^{(u)}_{\textup{III}}$ in \cite{partII}, when
the polynomial remote field of generic order is restricted  to the zero-order case (uniform Mode III).

\section{Full-field solution}

The full-field solution is presented for an infinite plane containing a single void and subject to torsion.
The void, with size defined by the smallest enclosing circle  of radius $a$, is considered
with different shape (the symbol embedded within square parentheses is used as apex to distinguish the respective solution):
\begin{itemize}
\item elliptical void $\left[\dellipse\right]$;
\item $n$-cusped hypocycloidal shaped void $\left[\text{\tiny\ding{71}}\right]$;
\item $n$-pointed isotoxal star-shaped polygonal void $\left[{\text{\tiny\ding{85}}}\right]$ (useful also to investigate the  particular cases of
$n$-pointed regular polygonal void $\left[\diamond\right]$, $n$-pointed regular star polygonal void $\left[{\text{\tiny\ding{73}}}\right]$,
$n$- pointed star-shaped cracks $\left[\ast\right]$, and the limit case of crack $\left[-\right]$).
\end{itemize}
Taking the centroid of the void inclusion as the origin of the reference system $\widehat{x}_1-\widehat{x}_2$,
with the axis $\widehat{x}_1$ aligned with the major axis (in the case of ellipse)
or with the bisector line of the inclusion vertex (in all the other cases), from eqn (\ref{xstarnostar}) the conformal mapping is given by
\beq\label{conf0}
\omega(\zeta)=\Upsilon a e^{\i \beta}+\, e^{\i \alpha}\widehat{\omega}(\zeta),
\eeq
where the function $\widehat{\omega}(\zeta)$ is the conformal mapping of the relevant void shape into the physical variable $\widehat{z}$ defined in
the system $\widehat{x}_1-\widehat{x}_2$,
\beq
\widehat{z}=\widehat{\omega}(\zeta),
\eeq
with $\widehat{z}=\widehat{x}_1+\i\widehat{x}_2$.
With reference to the conformal mapping (\ref{conf0}) particularized to a specific void shape,
evaluating the contour integral (\ref{eq_Schwarz_poisson_formula_final}) through Cauchy's integral formula (\ref{cauchycauchy}) provides the complex potential $g(\zeta)$,
showing the only dependence on the void  distance parameter $\Upsilon$ and the angular difference $(\alpha-\beta)$.

\subsection{Void with elliptical cross section}\label{ellipsesection}

The conformal mapping for an ellipse with major semi-axis  $a$ and minor semi-axis $\Lambda a$ (with $\Lambda\in[0,1]$), respectively parallel to the axes $\widehat{x}_1$ and $\widehat{x}_2$, is given by
\beq\label{confellipse}
\widehat{\omega}^{\left[\dellipse\right]}(\zeta)=a \frac{1+\Lambda}{2}\left(\frac{1}{\zeta}+\frac{1-\Lambda}{1+\Lambda} \zeta\right),
\eeq
where the apex $\left[\dellipse\right]$ refers to the elliptical hole problem. By applying Cauchy's integral formula (\ref{cauchycauchy}),
the complex potential $g^{\left[\dellipse\right]}(\zeta)$ (\ref{eq_Schwarz_poisson_formula_final}) follows as
\beq
g^{\left[\dellipse\right]}(\zeta)=\i a^2\left\{\,\frac{\left(1+\Lambda\right)^2}{4} \left[\frac{1-\Lambda}{1+\Lambda}\zeta^2+\left(\frac{1-\Lambda}{1+\Lambda}\right)^2+1\right]+\Upsilon\frac{1+\Lambda}{2}
\left( \frac{1-\Lambda}{1+\Lambda} e^{\i (\alpha-\beta)}+e^{-\i (\alpha-\beta)}\right)\zeta+\Upsilon^2\right\},
\eeq
and the shear stress components $\tau_{13}$ and $\tau_{23}$ can be therefore evaluated through eqn (\ref{eq_Stress_field_torsion_final}).

In order to analyze the stress state along the boundary of the elliptical void, the shear stress
modulus $\tau$ (\ref{eq_shear_stress}) is  evaluated along the unit disk boundary defined by $\zeta=e^{\i \eta}$ (being $\eta$ the  counter-clockwise angle within the conformal plane, so that if $\eta=0$ then $\zeta=1$) as
\beq\label{tauineta}
 \tau(\eta)=  \mu \Theta a  \left|
 \frac{\left[2+\left(1-\Lambda^2\right) \cos (2 \eta)\right]+2 \Upsilon (1+\Lambda) \cos (\eta -\alpha )}
 {1+\Lambda^2-\left(1-\Lambda^2\right) \cos 2 \eta }
 \right|
 \sqrt{ \sin ^2\eta+\Lambda^2 \cos ^2\eta}.
\eeq
Considering that the ellipse boundary can be parameterized through the radial distance $d(\chi)=\delta(\chi) a$ from the ellipse center
 as a function of the (anti-clockwise) polar angle $\chi$ (with $\chi=0$ for the point $\widehat{x}_1=a$, $\widehat{x}_2=0$ along the ellipse) and
\beq\label{distanzaellisse}
\delta(\chi)= \frac{\Lambda}{\sqrt{\Lambda^2 \cos^2\chi+\sin^2\chi}}
\eeq
the relations connecting the angle $\eta$ within the conformal plane to the
angle $\chi$ within the physical plane can be obtained from  the conformal mapping (\ref{confellipse}) as
\beq\label{cicciapuppa}
\sin\eta=\frac{\tan\chi}{\sqrt{\Lambda^2+\tan^2\chi}}\text{sign}{(\cos\chi)},\qquad
\cos\eta=\frac{\Lambda}{\sqrt{\Lambda^2+\tan^2\chi}}\text{sign}{(\cos\chi)}.
\eeq
By means of the trigonometrical relations (\ref{cicciapuppa}), the modulus of the shear stress (\ref{tauineta}) can be rewritten as the following function of the polar physical angle $\chi$,
\beq\label{tauinchi}
\begin{array}{lll}
\tau(\chi)=&\ds \mu\,\Theta a  \sqrt{\frac{\left(\Lambda ^2-1\right) \cos(2 \chi )+\Lambda ^2+1}{\left(\Lambda^4-1\right) \cos(2 \chi )+\Lambda^4+1}}\times\\
&\ds\times\left|\frac{\left(\Lambda ^2-1\right) \left(\Lambda^2-\tan ^2\chi\right)}{2\left(\Lambda ^2+\tan ^2\chi\right)}-
\frac{\Upsilon  (\Lambda +1) \text{sign}(\cos\chi )\left[\Lambda  \cos (\alpha-\beta)+\tan \chi \sin (\alpha -\beta )\right]}{\sqrt{\Lambda ^2+\tan ^2\chi}}
-\Lambda\right|.
\end{array}
\eeq
On the other hand, the  modulus of the unperturbed shear stress along the elliptical void boundary is
\beq
\tau^{\infty}(\chi)=\mu\,\Theta\,a \sqrt{\Upsilon^2+\delta^2(\chi)+2 \Upsilon \delta(\chi) \cos(\chi+\alpha-\beta)},
\eeq
which considering equation (\ref{distanzaellisse}) defining the radial distance parameter $\delta(\chi)$ reduces to
\beq\label{tauinfty}
\tau^{\infty}(\chi)=   \mu \Theta a  \sqrt{\Upsilon ^2+\frac{\Lambda ^2 \left(1+\tan ^2\chi\right)}{\Lambda ^2+\tan ^2\chi}
+\frac{2 \Upsilon  \Lambda  \text{sign}(\cos \chi) \left[\cos (\alpha-\beta)-\tan \chi  \sin (\alpha-\beta)\right]}{\sqrt{\Lambda ^2+\tan ^2(\chi )}}}.
\eeq

\subsection{$n$-cusped hypocycloidal shaped void}

The conformal mapping for a $n$-cusped hypocycloidal shaped void, inscribed in a circle of radius $a$ and having  a cusp at the coordinate $\widehat{x}_1=a$ and $\widehat{x}_2=0$,
is provided by 
\beq
\label{eq_map_ipocycloide}
\widehat{\omega}^{\left[\text{\tiny\ding{71}}\right]} (\zeta)=a \, \Omega^{\left[\text{\tiny\ding{71}}\right]}\left(\frac{1}{\zeta }+\frac{1}{n-1}\zeta^{n-1}\right),
\eeq
where (the apex ${\left[\text{\tiny\ding{71}}\right]}$ denotes the reference to the hypocycloidal void problem, and)
$\Omega^{\left[\text{\tiny\ding{71}}\right]}$ is a scaling factor depending on the number $n$ of cusps,
\beq
\Omega^{\left[\text{\tiny\ding{71}}\right]}(n)=\frac{n-1}{n} \in \left[\frac{1}{2},1\right).
\eeq
The complex potential for the torsion problem is computed in this case as
\beq
\label{eq_potenziale_ipocycloide}
g^{\left[\text{\tiny\ding{71}}\right]}(\zeta)=\i \,a^2\left[ \left(\Omega^{\left[\text{\tiny\ding{71}}\right]}\right)^2 \left(\frac{\zeta^n}{n-1}+\frac{n^2-2n+2}{(n-1)^2}\right)
+
\Upsilon\Omega^{\left[\text{\tiny\ding{71}}\right]} \left(\frac{\zeta^{n-1}}{n-1}e^{\i(\alpha-\beta)}+\zeta e^{-\i(\alpha-\beta)}\right)
+\Upsilon^2\right].
\eeq

Using the complex potential, eqn (\ref{eq_potenziale_ipocycloide}),
and the derivative of the conformal mapping (\ref{eq_map_ipocycloide}),
the stress components $\widehat{\tau}_{13}^{\left[\text{\tiny\ding{71}}\right]}$ and $\widehat{\tau}_{23}^{\left[\text{\tiny\ding{71}}\right]}$
in the transformed plane can be obtained from equation (\ref{eq_Stress_field_torsion_final}) as
\begin{dmath}
\label{tauconf_ipo_generale2}
\widehat{\tau}_{13}^{\left[\text{\tiny\ding{71}}\right]}-\i \widehat{\tau}_{23}^{\left[\text{\tiny\ding{71}}\right]}=\i \mu \Theta a
\left\{ \Omega^{\left[\text{\tiny\ding{71}}\right]} \left(\frac{n}{n-1}\frac{\zeta^{n+1}}{\zeta^{n}-1}-\frac{1}{\overline{\zeta}}-\frac{1}{n-1}\overline{\zeta\,}^{\,\,n-1}\right)
+\Upsilon e^{\i(\alpha-\beta)}\frac{\zeta^2+1}{\zeta^{n}-1}\right\},
\end{dmath}
highlighting the uncoupling in the stress field between the two \lq simple' remote loading conditions $(t)$ and $(u)$, as expected from the superposition principle due to the problem linearity
according to Sect. \ref{loaddec}.

The asymptotic representation is now derived for the stress field around the neighborhood of the cusp at $\widehat{x}_1 = a$ and $\widehat{x}_2=0$.  
According to Sect. \ref{stresssing}, with reference to  the radial distance $\rho$ from the inclusion cusp  and to
the counter-clockwise angle $\vartheta$ from  the $x_1$ axis, the complex vectors $\Delta z$ and $\Delta\zeta$ can be introduced as
\beq\label{zetazitaas0}
\Delta z=\Delta x_1+\i \Delta x_2=\rho e^{\i\vartheta},
\qquad
\Delta\zeta=\zeta-1,
\eeq
which are connected each other through the conformal mapping
\beq\label{zetazitaas}
z=\omega(1+\Delta\zeta)= a\left(\Upsilon e^{\i\beta}+ e^{\i\alpha}\right)+ \Delta z.
\eeq
Considering infinitesimal radial distance $\rho$ (so that $|\Delta z|\rightarrow 0$ and $|\Delta\zeta|\rightarrow 0$),
the inverse relation for the conformal mapping (\ref{eq_map_ipocycloide}) can be asymptotically obtained around the unit value for the complex variable $\zeta$ as
\begin{dmath}
\label{eq_inversa_ipocyc}
\Delta\zeta \simeq -\ds\sqrt{\frac{2\rho}{a (n-1)}}\, e^{\i \frac{\vartheta-\alpha}{2}},
\end{dmath}
which inserted in the stress field (\ref{tauconf_ipo_generale2}) leads to the following (square root singular) asymptotic expression
\beq
\label{asym_poly_z_star_generale_10}
\left[
\begin{array}{lll}
 \widehat\tau_{1 3}^{\left[\text{\tiny\ding{71}}\right]}\\
   \widehat\tau_{2 3}^{\left[\text{\tiny\ding{71}}\right]}
\end{array}
\right]
\simeq \ds \frac{\mu \Theta a}{n} \sqrt{\frac{a(n-1)}{2\rho}} \left[1+2 \Upsilon \cos \left(\alpha-\beta\right)\right]
\left[
   \begin{array}{lll}
   -\sin\left(\frac{\vartheta-\alpha}{2}\right)\\
   \cos\left(\frac{\vartheta-\alpha}{2}\right)
   \end{array}
   \right].
\eeq

Recalling the Stress Intensity Factor definition, eqn (\ref{eq_def_nsif}), the closed-form expression for $K^{\left[\text{\tiny\ding{71}}\right]}
_{\text{III}}$  can be obtained from the asymptotic fields (\ref{asym_poly_z_star_generale_10}) as
\begin{dmath}
K^{\left[\text{\tiny\ding{71}}\right]}
_{\text{III}}=\mu \Theta a \frac{\sqrt{(n-1) \pi a }}{n}\left[1+ 2\Upsilon \cos \left(\alpha-\beta\right)\right].
\end{dmath}
The asymptotic behaviour (\ref{asym_poly_z_star_generale_10}) can be extended to describe the response around the generic $k$-th cusp (with $k=1,\cdots,n$)
by substituting the angle $\alpha$ with the angle $\alpha_k$ defined as
\beq
\alpha_k=\alpha+\frac{2\pi(k-1)}{n},
\eeq
from which the  Stress Intensity Factor $K^{\left[\text{\tiny\ding{71}}\right]}
_{\text{III}}(k)$ ruling the singularity raised at the $k$-th cusp (with $k=1,\cdots,n$) of the hypocycloidal void follows as
\beq
\label{eq_sif_hypo}
K^{\left[\text{\tiny\ding{71}}\right]}
_{\text{III}}(k)=\mu \Theta a\frac{\sqrt{ (n-1) \pi a}}{n}\left[1+ 2 \Upsilon \cos \left(\alpha-\beta+\frac{2\pi(k-1)}{n}\right)\right].
\eeq

\subsection{Isotoxal star-shaped polygonal void }

The conformal mapping from the unit disk onto the plane containing a polygonal inclusion is provided by the Schwarz-Christoffel formula.
In the case of $n$-pointed (non-intersecting) isotoxal star-shaped polygonal voids (having a semi-angle at the isotoxal-points $\xi \pi$, with $\xi\in[0,(n-2)/2n]$),
 the conformal mapping  $\widehat{\omega}^{\left[\text{\ding{85}}\right]}(\zeta)$ is given by
 \beq\label{omegahatisotoxal}
 \widehat{\omega}^{\left[\text{\ding{85}}\right]}(\zeta)=\ds a \Omega^{\left[\text{\ding{85}}\right]} \int^{\zeta}\left[\frac{1}{\varsigma^2}\prod^{n-1}_{j=0}\left(1- \varsigma e^{-\i\frac{2 k\pi}{n}}\right)^{1-2\xi}
 \prod^{n}_{j=1}\left(1- \varsigma e^{-\i\frac{(2 k-1)\pi}{n}}\right)^{2(\xi+1/n)-1}\right]\,\mathrm{d}\varsigma,
 \eeq
where (the apex ${\left[\text{\ding{85}}\right]}$ refers to the isotoxal void problem,)
$a$ is the radius of the circle inscribing the void and $\Omega^{\left[\text{\ding{85}}\right]}$ is a real parameter depending on the shape of the void as follows
\beq
\label{eq_shape_parameter_poligoni_regolari_stella_generale}
\Omega^{\left[\text{\ding{85}}\right]}(n,\xi)= \frac{1}{\sqrt[n]{4}}\dfrac{\Gamma\left(1-\frac{1}{n}-\xi\right)}
{\Gamma\left(\frac{n-1}{n}\right)\Gamma\left(1-\xi\right)} \in \left[\frac{1}{2},1\right),
\eeq
with the symbol $\Gamma(\cdot)$ standing for the Euler gamma function defined via the following convergent improper integral
\beq
\Gamma(x)=\int_{0}^{\infty} y^{x-1} e^{-y} dy.
\eeq

Recalling that $e^{-\i\frac{2 j\pi}{n}}$ and $e^{-\i\frac{(2 j-1)\pi}{n}}$ are respectively the $n$-th roots of the positive and negative unity, the following properties
\beq
\prod^{n-1}_{j=0}\left(1- \varsigma e^{-\i\frac{2 j\pi}{n}}\right)=1-\varsigma^n, \qquad
 \prod^{n}_{j=1}\left(1- \varsigma e^{-\i\frac{(2 j-1)\pi}{n}}\right)=1+\varsigma^n,
\eeq
can be used to simplify the conformal mapping (\ref{omegahatisotoxal}) as
\beq
\label{eq_sc_integral_ext_ext_regular_polygon_diffrential_form_simple_stella_generale}
\widehat{\omega}^{\left[\text{\ding{85}}\right]}(\zeta)=-\ds a \Omega^{\left[\text{\ding{85}}\right]}\,
 \int^{\zeta}_1
 \frac{\left(1-\varsigma^{n}\right)^{1-2\xi}\left(\varsigma^{n}+1\right)^{2\left(\xi+\frac{1}{n}\right)-1}}{\varsigma^{2}}\mathrm{d}\varsigma,
\eeq
an equation that can be reduced by introducing the Appell hypergeometric function $F_1$,
\beq
\widehat{\omega}^{\left[\text{\ding{85}}\right]}(\zeta)=\frac{a \Omega^{\left[\text{\ding{85}}\right]}  F_1\left(-\frac{1}{n};2 \xi-1,1-2\xi -\frac{2}{n};1-\frac{1}{n};\zeta ^n,-\zeta ^n\right)}{\zeta }.
\eeq
By definition, the Appell hypergeometric function $F_1$ can be expressed by a series, so that the conformal mapping $\widehat{\omega}^{\left[\text{\ding{85}}\right]}(\zeta)$
is equivalently given by the following Laurent series
\beq
\label{conformal_map_approx}
\widehat\omega^{\left[\text{\ding{85}}\right]}(\zeta) = a \Omega^{\left[\text{\ding{85}}\right]} \sum^{\infty}_{j=0}  d_{j}^{\left[\text{\ding{85}}\right]}(\xi,n) \,\zeta^{j n-1}\,,
\eeq
where the real constants $d_j^{\left[\text{\ding{85}}\right]}(\xi,n)$ are dependent on the angle ratio $\xi$ and the point number $n$ and are defined as
\begin{equation}
\lb{checazzo}
d_{j}^{\left[\text{\ding{85}}\right]}(\xi,n)=\frac{1}{1-j n} \sum_{k=0}^{j}
\frac{(-1)^{j-k}}{k!(j-k)!}
\frac{\Gamma\left(1-\frac{2}{n}-2\xi+j-k\right)\Gamma\left(-1+2\xi+k\right)}
{\Gamma\left(1-\frac{2}{n}-2\xi \right)\Gamma\left(-1+2\xi\right)} ,
\end{equation}
and satisfy the following property

\beq\label{propincredibile}
\sum_{j=1}^{\infty} (n j-1) d_j^{\left[\text{\ding{85}}\right]}(\xi,n)=1, \qquad \forall\, n.
\eeq

In order to compute the complex potential $g^{\left[\text{\ding{85}}\right]}(\zeta)$, it is fundamental to  evaluate first the quantity $\omega^{\left[\text{\ding{85}}\right]}(\sigma)
\overline{\omega{\left[\text{\ding{85}}\right]}(\sigma)}$ which, considering
the Laurent series for the conformal mapping (\ref{conformal_map_approx}), is given by
\beq
\omega^{\left[\text{\ding{85}}\right]}(\sigma)\overline{\omega^{\left[\text{\ding{85}}\right]}(\sigma)}=a^2\left\{
\left(\Omega^{\left[\text{\ding{85}}\right]}\right)^2
\,\sum_{j=0}^{\infty}\sum_{l=0}^{\infty} d_j^{\left[\text{\ding{85}}\right]} d_l^{\left[\text{\ding{85}}\right]}\sigma^{n(j-l)}+ \Upsilon \Omega^{\left[\text{\ding{85}}\right]}  \left[
 \, \sum_{j=0}^{\infty} d_j^{\left[\text{\ding{85}}\right]} \, \left(e^{\i (\alpha-\beta)}\sigma^{n j-1}
+ \, e^{-\i (\alpha-\beta)}  \sigma^{1-n j}\right)
\right]+\Upsilon^2\right\}.
\eeq
and the substitution of this quantity  in formula (\ref{eq_Schwarz_poisson_formula_final}) provides, after mathematical manipulation, the complex potential
\beq
g^{\left[\text{\ding{85}}\right]}(\zeta)=\i\,a^2\left[\left(\Omega^{\left[\text{\ding{85}}\right]}\right)^2
\,\sum_{j=0}^{\infty}\sum_{l=0}^{j} d_j^{\left[\text{\ding{85}}\right]} d_l^{\left[\text{\ding{85}}\right]}\zeta^{n(j-l)}+ \Upsilon \Omega^{\left[\text{\ding{85}}\right]}  \left(
 \, e^{\i (\alpha-\beta)}\sum_{j=1}^{\infty} d_j^{\left[\text{\ding{85}}\right]} \, \zeta^{n j-1}
+ \, e^{-\i (\alpha-\beta)} \zeta
\right)+\Upsilon^2\right].
\eeq

The shear stress field in the conformal plane is given by
\begin{dmath}
\label{tauconf_star_isotoxal}
\widehat{\tau}_{13}^{\left[\text{\ding{85}}\right]}-\i \widehat{\tau}_{23}^{\left[\text{\ding{85}}\right]}=
-\i \mu \Theta a \left\{ \Omega^{\left[\text{\ding{85}}\right]}\left[\ds\frac{\ds n\sum_{j=0}^{\infty}\sum_{l=0}^{j} (j-l) d_j^{\left[\text{\ding{85}}\right]} d_l^{\left[\text{\ding{85}}\right]}\zeta^{n(j-l)+1}}
{\left(1-\zeta^{n}\right)^{1-2\xi}\left(\zeta^{n}+1\right)^{2\left(\xi+\frac{1}{n}\right)-1}}+
\sum_{j=0}^{\infty} d_j^{\left[\text{\ding{85}}\right]}\overline{\zeta}^{j n- 1}
\right]
+\Upsilon\left[\ds
\frac{e^{\i (\alpha-\beta)}\ds\sum_{j=1}^{\infty} (n j-1) d_j^{\left[\text{\ding{85}}\right]} \zeta^{n\, j}
+\zeta^2 e^{-\i (\alpha-\beta)}}{\left(1-\zeta^{n}\right)^{1-2\xi}\left(\zeta^{n}+1\right)^{2\left(\xi+\frac{1}{n}\right)-1}}
+e^{\i (\alpha-\beta)}\right]\right\}.
\end{dmath}

Integrating the expansion around the unit value of the transformed variable, $\zeta=1+\Delta \zeta$,
of the first derivative of the conformal mapping (\ref{eq_sc_integral_ext_ext_regular_polygon_diffrential_form_simple_stella_generale}),
the expansion of the conformal mapping can be obtained as
\beq
\widehat{\omega}^{\left[\text{\ding{85}}\right]}(\zeta)=-a \Omega^{\left[\text{\ding{85}}\right]}\,
  \frac{\left(1-\zeta^{n}\right)^{1-2\xi}\left(\zeta^{n}+1\right)^{2\left(\xi+\frac{1}{n}\right)-1}}{\zeta^{2}},
\eeq
and the comparison with eqn (\ref{eq_inversa_ipocyc}) provides the following
asymptotic relation between the physical coordinate $\Delta z$ and its conformal counterpart $\Delta \zeta$
\beq
\label{eq_relationship_between_two_planes_regolare_star_generale}
\Delta \zeta \simeq \ds -
2^{\frac{n(1-\xi)-1}{n(1-\xi)}
}
n^{\frac{2\xi-1}{2(1-\xi)}}\left(\frac{1-\xi}{a \,\Omega^{\left[\text{\ding{85}}\right]}(n,\xi)}\rho \right)^{\frac{1}{2(1-\xi)}} e^{\i\frac{\vartheta-\alpha}{2(1-\xi)}}.
\eeq
Using the asymptotic inverse relation (\ref{eq_relationship_between_two_planes_regolare_star_generale}) in the
stress field (\ref{tauconf_star_isotoxal}) and the property (\ref{propincredibile})
 provides the following leading order term for the shear stress
\begin{dmath}
\label{tauconf_star_isotoxal_asym}
\left[
\begin{array}{lll}
\widehat{\tau}_{13}^{\left[\text{\ding{85}}\right]}\\
\widehat{\tau}_{23}^{\left[\text{\ding{85}}\right]}
\end{array}
\right]
\simeq
\frac{\mu \Theta a}{2^{\frac{1}{n(1-\xi)}}}\left(\frac{ a \Omega^{\left[\text{\ding{85}}\right]}}{n (1-\xi) \rho}\right)^{\frac{1-2\xi}{2(1-\xi)}}\left(n   \Omega^{\left[\text{\ding{85}}\right]}\ds
\ds\sum_{j=0}^{\infty}\sum_{l=0}^{j} (j-l) d_j^{\left[\text{\ding{85}}\right]} d_l^{\left[\text{\ding{85}}\right]}
+2 \Upsilon \cos \left(\alpha-\beta\right)\right)
\left[
   \begin{array}{lll}
   -\sin\left(\frac{1-2\xi}{2(1-\xi)}(\vartheta-\alpha)\right)\\
   \cos\left(\frac{1-2\xi}{2(1-\xi)}(\vartheta-\alpha)\right)
   \end{array}
   \right].
\end{dmath}

Recalling the definition (\ref{eq_def_nsif}), the Notch Stress Intensity Factor at the $k$-th point of the isotoxal polygonal void results
\begin{dmath}
\label{Kisotoxal}
K^{\left[\text{\ding{85}}\right]}_{\text{III}}(k)=2^{\frac{n(1-\xi)-2}{2n(1-\xi)}} \sqrt{\pi} \mu \Theta a
\left(\frac{ a \Omega^{\left[\text{\ding{85}}\right]}}{n (1-\xi)}\right)^{\frac{1-2\xi}{2(1-\xi)}}\left(n   \Omega^{\left[\text{\ding{85}}\right]}\ds
\ds\sum_{j=0}^{\infty}\sum_{l=0}^{j} (j-l) d_j^{\left[\text{\ding{85}}\right]} d_l^{\left[\text{\ding{85}}\right]}
+2 \Upsilon \cos \left(\alpha-\beta+\frac{2\pi(k-1)}{n}\right)\right).
\end{dmath}

The stress intensification (\ref{Kisotoxal}) is particularized to the speciale case of  $n$-sided regular polygonal voids ($\xi=1/2-1/n$),
$n$-pointed regular star polygonal voids ($\xi=1/2-2/n$), and $n$-pointed star-shaped cracks
($\xi=0$).

\subsubsection{$n$-sided regular polygonal void}

Assuming the semi-angle ratio as $\xi=1/2-1/n$, the isotoxal inclusion reduces to a $n$-sided regular polygonal void
and the scale factor $\Omega^{\left[\text{\ding{85}}\right]}$, eqn (\ref{eq_shape_parameter_poligoni_regolari_stella_generale}), simplifies as
\beq
\Omega^{\left[\diamond\right]}=\frac{\sqrt{\pi} }{\sqrt[n]{4}\,\,\Gamma\left(\frac{1}{2}+\frac{1}{n}\right)\Gamma\left(1-\frac{1}{n}\right)},
\eeq
while the real constants $d_j^{\left[\text{\ding{85}}\right]}$, eqn (\ref{checazzo}), as
\beq
d_j^{\left[\diamond\right]}=\frac{\Gamma\left(j-\frac{2}{n}\right)}{j!(1-j n)\Gamma\left(-\frac{2}{n}\right)},
\eeq
where the apex ${\left[\diamond\right]}$ denotes the reference to the $n$-sided regular polygonal void problem. For this void shape, the NSIF (\ref{Kisotoxal}) reduces to the following
expression
\begin{dmath}\label{NSIFpolygon}
K^{\left[\diamond\right]}_{\text{III}}(k)= \sqrt{2\pi} \mu \Theta a
\left(\frac{ a \Omega^{\left[\diamond\right]}}{n +2}\right)^{\frac{2}{n+2}}\left(
n   \Omega^{\left[\diamond\right]}\ds
\ds\sum_{j=0}^{\infty}\sum_{l=0}^{j} (j-l) d_j^{\left[\diamond\right]} d_l^{\left[\diamond\right]}
+2 \Upsilon \cos \left(\alpha-\beta+\frac{2\pi(k-1)}{n}\right)\right).
\end{dmath}

\subsubsection{$n$-pointed regular star polygonal void}

The isotoxal inclusion reduces to a $n$-pointed regular star polygonal void (with value 2 of starriness and $n\geq 5$, see \cite{partI}) taking
the semi-angle ratio as $\xi=1/2-2/n$. Under this geometry restriction,
the scale factor $\Omega^{\left[\text{\ding{85}}\right]}$, eqn (\ref{eq_shape_parameter_poligoni_regolari_stella_generale}), becomes
\beq
\Omega^{\left[\text{\ding{73}}\right]}(n)=
\frac{\sin\left(\frac{\pi}{n}\right)}{\pi}\frac{\Gamma\left(\frac{2}{n}\right)^2 }{\Gamma\left(\frac{4}{n}\right)},
\eeq
and the real constants $d_j^{\left[\text{\ding{85}}\right]}$, eqn (\ref{checazzo}), as
\beq
d_j^{\left[\text{\ding{73}}\right]}=\frac{1}{1-j n} \sum_{k=0}^{j} \frac{(-1)^{j-k}}{k!(j-k)!}
\frac{\Gamma\left(\frac{2}{n}+j-k\right)
\Gamma\left(-\frac{4}{n}+k\right)}{\Gamma\left(\frac{2}{n}\right) \Gamma\left(-\frac{4}{n} \right)},
\eeq
where the apex $\left[\text{\ding{73}}\right]$ denotes the reference to the $n$-pointed star polygonal void problem.
In this case, the NSIF (\ref{Kisotoxal}) also simplifies  as
\begin{dmath}\label{NSIFstar}
K^{\left[\text{\ding{73}}\right]}_{\text{III}}(k)=2^{\frac{n+8}{2(n+4)}} \sqrt{\pi} \mu \Theta a
\left(\frac{ a \Omega^{\left[\text{\ding{73}}\right]}}{n+4}\right)^{\frac{4}{n+4}}\left(
n   \Omega^{\left[\text{\ding{73}}\right]}\ds
\ds\sum_{j=0}^{\infty}\sum_{l=0}^{j} (j-l) d_j^{\left[\text{\ding{73}}\right]} d_l^{\left[\text{\ding{73}}\right]}
+2 \Upsilon \cos \left(\alpha-\beta+\frac{2\pi(k-1)}{n}\right)\right).
\end{dmath}

\subsubsection{$n$-pointed star-shaped crack}

In the case of $n$-pointed star-shaped cracks ($\xi=0$), the integral  in Schwarz-Christoffel formula (\ref{eq_sc_integral_ext_ext_regular_polygon_diffrential_form_simple_stella_generale})
can be analytically evaluated, so that the conformal mapping reduces to the following simple expression
\begin{dmath}\label{confssc}
\widehat{\omega}^{\left[\ast\right]} (\zeta)= a \, \Omega^{\left[\ast\right]}\frac{\left(1+\zeta^n\right)^{\frac{2}{n}}}{\zeta},
\end{dmath}
where
\beq
\label{eq_shape_parameter_poligoni_regolari_stella_generale_crack}
\Omega^{\left[\ast\right]}(n,\xi)= \frac{1}{\sqrt[n]{4}}\dfrac{\Gamma\left(1-\frac{1}{n}-\xi\right)}
{\Gamma\left(\frac{n-1}{n}\right)\Gamma\left(1-\xi\right)} \in \left[\frac{1}{2},1\right),
\eeq
and the apex $\left[\ast\right]$ denotes the reference to the star-shaped crack problem. The shear stress field can be evaluated from eqn (\ref{eq_Stress_field_torsion_final})
 as a function of the transformed complex variable $\zeta$ as
\begin{dmath}
\label{tauconf_star_crack_generale}
\widehat{\tau}_{13}^{\left[\ast\right]}-\i \widehat{\tau}_{23}^{\left[\ast\right]}=
\i \mu \Theta a\left\{ \Omega^{\left[\ast\right]}\left[\ds\frac{\ds n\sum_{j=0}^{\infty}\sum_{l=0}^{j} (j-l) d_j^{\left[\ast\right]} d_l^{\left[\ast\right]}\zeta^{n(j-l)+1}}{\left(\zeta^n-1\right)\left(\zeta^n+1\right)^{\frac{2}{n}-1}}
-\frac{\overline{\left(\zeta^n+1\right)^{\frac{2}{n}}}}{\overline{\zeta}}\right]
+\Upsilon\left[\ds
\frac{e^{\i (\alpha-\beta)}\ds\sum_{j=1}^{\infty} (n j-1) d_j^{\left[\ast\right]} \zeta^{n\, j}
+\zeta^2 e^{-\i (\alpha-\beta)}}{\left(\zeta^n-1\right)\left(\zeta^n+1\right)^{\frac{2}{n}-1}}
-e^{\i (\alpha-\beta)}\right]\right\}.
\end{dmath}
Expanding  the conformal mapping (\ref{confssc}) around the tip of the isotoxal star-shaped crack, eqn (\ref{zetazitaas}), provides the inverse relation between the infinitesimal complex quantities $\Delta \zeta$ and $\Delta z=\rho e^{\i \vartheta}$ as
\beq
\label{eq_inversa_star_crac}
\Delta\zeta \simeq - \ds 2 \sqrt{\frac{\rho}{n a}  }\,\, e^{\frac{\i(\vartheta-\alpha)}{2}},
\eeq
which used in the shear stress field (\ref{tauconf_star_crack_generale}) provides the following asymptotic expression
\begin{dmath}
\label{tauconf_star_crack_asym}\left[
\begin{array}{lll}
\widehat{\tau}_{13}^{\left[\ast\right]}\\
\widehat{\tau}_{23}^{\left[\ast\right]}
\end{array}
\right]
\approx
\frac{\mu \Theta a}{\sqrt[n]{4}}\sqrt{\frac{a}{n \rho}} \left(\frac{n}{\sqrt[n]{4}}\ds
\ds\sum_{j=0}^{\infty}\sum_{l=0}^{j} (j-l) d_j^{\left[\ast\right]} d_l^{\left[\ast\right]}
+2 \Upsilon \cos \left(\alpha-\beta\right)\right)
\left[
   \begin{array}{lll}
   -\sin\left(\frac{\vartheta-\alpha}{2}\right)\\
   \cos\left(\frac{\vartheta-\alpha}{2}\right)
   \end{array}
   \right],
\end{dmath}
obtained exploiting  property expressed by eqn (\ref{propincredibile}).

The closed form expression for the Stress Intensity Factor $K^{\left[\ast\right]}_{\text{III}}(k)$  related to the $k$-th point($k=1,\cdots,n$) of a
 $n$-pointed isotoxal star-shaped crack can be evaluated
considering definition (\ref{eq_def_nsif}) and the asymptotic fields (\ref{tauconf_star_crack_asym}) as
\begin{equation}
\label{k3ssc}
K^{\left[\ast\right]}_{\text{III}}(k)=\ds 2 ^{\frac{n-4}{2n}}\mu \Theta a\sqrt{\frac{\pi a}{n}} \left[\frac{n }{\sqrt[n]{4}}\ds
\ds\sum_{j=0}^{\infty}\sum_{l=0}^{j} (j-l) d_j^{\left[\ast\right]} d_l^{\left[\ast\right]}
+2 \Upsilon \cos \left(\alpha-\beta+\frac{2\pi(k-1)}{n}\right)\right],
\end{equation}
which after mathematical manipulation can be rewritten as the following series
\begin{dmath}
\label{eq_sif_star_crack}
K^{\left[\ast\right]}_{\text{III}}(k)=\ds 2 ^{\frac{n-4}{2n}}\mu \Theta a\sqrt{\frac{\pi a}{n}} \left[\frac{ 2}{n} \left[\Gamma\left(\frac{2}{n}\right)\right]^2\ds
\ds\sum_{j=0}^{\infty}
\frac{ {}_2F_1\left[j,1+\frac{2}{n};2+j;-1\right]}{\Gamma\left(1+j\right)\Gamma\left(2+j\right)\Gamma\left(\frac{2}{n}-j\right)\Gamma\left(1-j+\frac{2}{n}\right)}
+2\Upsilon \cos \left(\alpha-\beta+\frac{2\pi(k-1)}{n}\right)\right],
\end{dmath}
which results to be more rapidly convergent than expression (\ref{k3ssc}). In equation (\ref{eq_sif_star_crack}) the
symbol ${}_2F_1$ stands for hyper-geometric function defined for $|z| < 1$ by the power series
\beq
{}_2F_1\left[a,b;c;z\right]= \sum_{j=0}^{\infty}\frac{\left(a\right)_j\left(b\right)_j}{\left(c\right)_j} \frac{z^j}{j!},
\eeq
where $\left(x\right)_j$ gives the Pochhammer symbol (rising factorial) given as
\beq
\left(x\right)_j=\frac{\Gamma(x+j)}{\Gamma(x)}=x(x+1)\cdots(x+j-1).
\eeq

\begin{figure}[h!]
  \begin{center}
\includegraphics[width=17 cm]{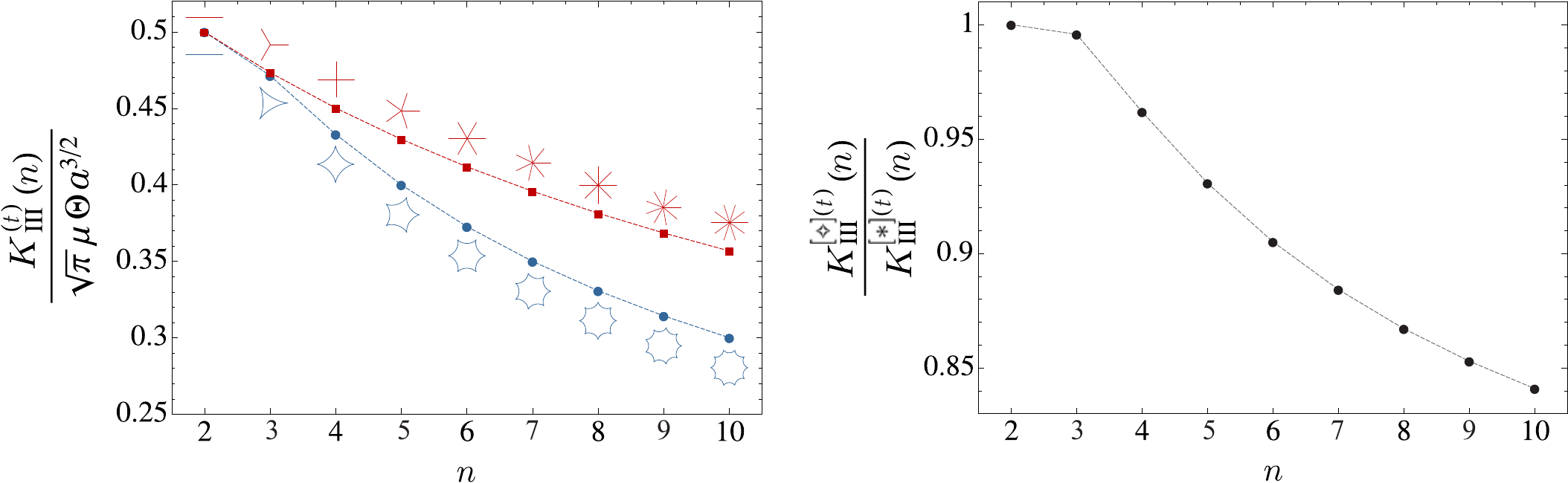}
\caption{
\small Stress Intensity Factors $K^{\left[\ast\right]}_{\text{III}}$ for $n$-pointed star-shaped cracks (red squares), eqn (\ref{eq_sif_star_crack}), and $K^{\left[\text{\tiny\ding{71}}\right]}_{\text{III}}$
for $n$-cusped  hypocycloidal shaped void
(blue circles), eqn (\ref{eq_sif_hypo}), at varying the number $n$ of tips/cusps in the case of a null radial distance for the void, $\Upsilon=0$, so that $K^{\left[\ast\right]}_{\text{III}}=K^{\left[\ast\right](t)}_{\text{III}}$
and $K^{\left[\text{\tiny\ding{71}}\right]}_{\text{III}}=K^{\left[\text{\tiny\ding{71}}\right](t)}_{\text{III}}$. (Left) Values normalized through division by $\sqrt\pi\mu \Theta a^{3/2} $ and
(right) ratio of the values corresponding to the two shapes with same $n$. The comparison shows that SIF for the star-shaped cracks is never smaller than that for the hypocycloidal void for the same $n$.
}
\label{crackvshypo}
 \end{center}
\end{figure}

With reference to \lq pure' torsion condition, $\Upsilon=0$, the Stress Intensity Factor $K^{\left[\ast\right](t)}_{\text{III}}$
for a $n$-pointed star-shaped crack, eqn (\ref{eq_sif_star_crack}),
is reported in Fig. \ref{crackvshypo}  at varying the value $n$ and compared with the corresponding quantity $K^{\left[\text{\tiny\ding{71}}\right](t)}_{\text{III}}$
for a $n$-cusped hypocycloidal shaped void, eqn (\ref{eq_sif_hypo}).
It is observed that the SIF in the former case is always greater than that in the latter, except for $n=2$ where the two corresponding values coincide because 
both inclusion shapes reduce to the \lq standard' crack geometry.

\paragraph{The \lq standard' crack.}
The solution in the case of a \lq standard' crack ($n=2$), denoted by the symbol $\left[-\right]$, is reported as a specialization of the solution for $n$-pointed star-shaped cracks.
In this case the conformal mapping $\omega(\zeta)$, eqns (\ref{confssc}) and (\ref{conf0}) , reduce to the simple expression
\beq
\omega^{\left[-\right]}(\zeta)=e^{\i \alpha}\frac{a}{2}\left(\frac{1}{\zeta}+ \zeta\right)+\Upsilon a e^{\i \beta}=z,
\eeq
of which inverse is explicitly given by
\beq\label{inversecrack}
\zeta=e^{-\i \alpha}\left(\frac{z}{a}-\Upsilon  e^{\i \beta}\right)-\sqrt{e^{-\i \alpha}\left(\frac{z}{a}-\Upsilon  e^{\i \beta}\right)+1} \sqrt{e^{-\i \alpha}\left(\frac{z}{a}-\Upsilon  e^{\i \beta}\right)-1}.
\eeq
The complex potential in the conformal plane can be obtained specializing
the solution for the elliptical void (in the limit of null $\Lambda$), or that for the $n$-pointed star-shaped inclusion
(taking $n=2$), as
\beq
g^{\left[-\right]}(\zeta)=\i a^2 \left[\frac{\zeta^2+2}{4} + \Upsilon \zeta \cos (\alpha-\beta) +\Upsilon^2\right],
\eeq
which, through the inverse relation (\ref{inversecrack}), can be also explicitly expressed in the physical plane as
\beq
G^{\left[-\right]}(\widehat{z})=\frac{\i }{4} \left\{a^2\bigg[1+4 \Upsilon\bigg(\Upsilon+  \cos (\beta -\alpha )\bigg)\bigg]-2 \widehat{z}\sqrt{\widehat{z}-a} \sqrt{\widehat{z}+a}+2 \widehat{z}^2\right\}.
\eeq

After derivation of the complex potential, the stress fields can be expressed in the $\widehat{x}_1-\widehat{x}_2$ system through the following formula
\beq
\widehat{\tau}_{13}^{\left[-\right]}(\widehat{z})-\i\widehat{\tau}^{\left[-\right]}_{23}(\widehat{z})=
-\frac{\i \mu\Theta }{2}      \left(\frac{\left(\widehat{z}-\sqrt{\widehat{z}-a} \sqrt{\widehat{z}+a}\right)^2}{\sqrt{\widehat{z}-a} \sqrt{\widehat{z}+a}}+2 \Upsilon a e^{-i (\beta -\alpha )}
+2 \overline{\widehat{z}}\right),
\eeq
which expanded at the tip with coordinate $\widehat{x}_1=a$ and $\widehat{x}_2=0$, so that for $\widehat{z}=a+\rho e^{\i (\vartheta-\alpha)}$ with small radial distance $\rho$, provides
the asymptotic expression of the shear stress field
\beq
\label{asym_crack2}
\left[
\begin{array}{lll}
 \widehat\tau_{1 3}^{\left[-\right]}\\
   \widehat\tau_{2 3}^{\left[-\right]}
\end{array}
\right]
\simeq \ds \frac{\mu \Theta a }{2} \sqrt{\frac{a}{2\rho}} \left[1+2 \Upsilon \cos \left(\alpha-\beta\right)\right]
\left[
   \begin{array}{lll}
   \ds-\sin\left(\frac{\vartheta-\alpha}{2}\right)\\[6mm]
   \ds\cos\left(\frac{\vartheta-\alpha}{2}\right)
   \end{array}
   \right],
\eeq
and the SIF for the $k$-th tip ($k=1,2$) is given by
\beq\label{sif_crack_n2}
K^{\left[-\right]}_{\text{III}}(k)=\mu \Theta a\frac{\sqrt{\pi a }}{2} \left[1+ 2\Upsilon \cos \left(\alpha-\beta+\pi(k-1)\right)\right].
\eeq

It is should be noted that the SIF given by eqn (\ref{sif_crack_n2}) in the particular case of $\Upsilon=0$ coincides with that obtained by Sih \cite{sih}
for a crack centered in an elliptical bar in the limit case of infinite cross section (except for the multiplication with $\sqrt{\pi}$ because of the slightly different
definition of SIF used therein).

\section{Discussion on the analytical results}

The non-singular (elliptical void) and singular (hypocycloidal and isotoxal voids) stress fields presented in the previous Section are analyzed with the aim to disclose the role of the void location on the stress intensification and the possibility of stress reduction.

\subsection{Stress concentration and stress annihilation along the elliptical void boundary}

The stress amplification due to the presence of the elliptical void can be evaluated through the Stress Concentration Factor (SCF). This parameter is defined as the ratio of modulus of the shear stress
in the presence of the void $\tau$, eqn (\ref{tauinchi}), and that in the case of the absence of the void (unperturbed field) $\tau^{\infty}$, eqn (\ref{tauinfty}),
\beq\label{scf}
\text{SCF}(\chi)=\frac{\tau(\chi)}{\tau^{\infty}(\chi)}.
\eeq
Because of the inherent positiveness of the shear stress modulus, eqn (\ref{eq_shear_stress}), SCF is a non-negative parameter showing the increase (or reduction) of
the stress state along the inclusion boundary by the presence of the void when SCF is greater (or smaller) than one. A null value for the stress concentration (SCF=0) may be verified for
some angular coordinate $\widetilde\chi$ when stress annihilation occurs at this point, $\tau(\widetilde\chi)=0$. The existence of such an angle is affected
by the parameters $\Upsilon$, $\Lambda$ and $(\alpha-\beta)$ and its value provided by the following condition
\begin{dmath}
 2\Upsilon (\Lambda +1) \text{sign}(\cos\widetilde{\chi} )
\left[\Lambda  \cos (\alpha-\beta)+\tan \widetilde{\chi} \sin (\alpha -\beta )\right] \sqrt{\Lambda ^2+\tan ^2\widetilde{\chi}}+2\Lambda \left(\Lambda ^2+\tan ^2\widetilde{\chi}\right)
-\left(\Lambda ^2-1\right) \left(\Lambda^2-\tan ^2\widetilde{\chi}\right)=0.
\end{dmath}
At varying of the parameters $\Lambda$, $\Upsilon$, and ($\alpha-\beta$), stress annihilation (SCF=0)
is numerically found to possibly occur  at (i.) four points, (ii.) two points, or (iii.) no point along the ellipse boundary.
Restricting the attention to the case of ellipse center coincident with the origin of the $x_1-x_2$ system ($\Upsilon=0$),
the expression (\ref{scf}) for the SCF reduces to
\beq\label{scfchizero}
\text{SCF}(\chi)=\frac{\left| \Lambda ^4-2 \Lambda ^3-2 \Lambda ^2-2 \Lambda +1 +(\Lambda +1)(\Lambda -1)^3 \cos (2 \chi )\right| }{2 \sqrt{2} \Lambda
\sqrt{\left(\Lambda ^4-1\right) \cos (2 \chi )+\Lambda ^4+1}},
\eeq
showing the independence of the angular difference $\alpha-\beta$ and that stress annihilation (SCF=0) occurs along the void boundary  (i.)
at four points when $\Lambda \in [0,\sqrt{2}-1)$, (ii.) at  two points when $\Lambda=\sqrt{2}- 1$,  and (iii.) at no point
when $\Lambda \in (\sqrt{2}-1,1]$.

\begin{figure}[h!]
  \begin{center}
\includegraphics[width=16 cm]{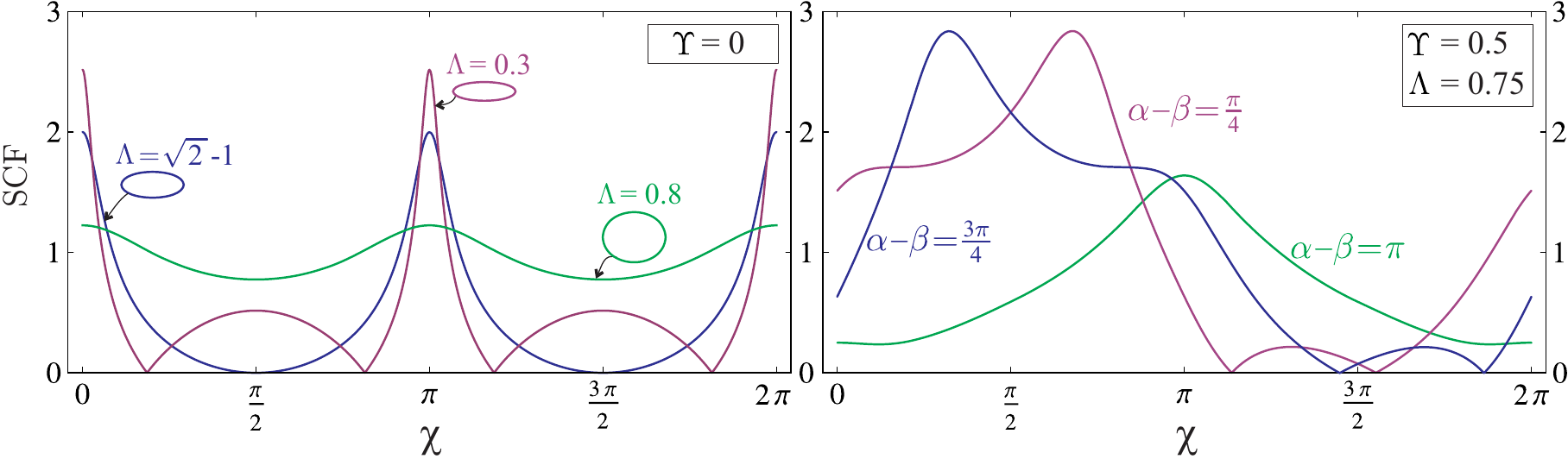}
\caption{
\small Stress Concentration Factor (SCF) along the void boundary displayed as a function of the polar angle $\chi$.
(Left) The case of elliptical void centered at the origin of the $x_1-x_2$ system $(\Upsilon=0)$ for three  values of $\Lambda=\left\{0.3,\sqrt{2}-1,0.8\right\}$.
(Right) The case of elliptical void with a distance $\Upsilon=0.5$ and $\Lambda=0.75$ for  three values of the  angle difference $(\alpha-\beta)=\left\{1/4,3/4,1\right\}\pi$.
Stress annihilation occurring at none, two and four points along the ellipse boundary can be observed at varying the geometry parameters.}
\label{fig_scf}
 \end{center}
\end{figure}

The possible occurrence of stress annihilation can be observed in Figure \ref{fig_scf}, where the cases $\Upsilon=0$ and $\Upsilon=0.5$ are reported
respectively on the left and on the right.
More specifically, in the case $\Upsilon=0$ (Fig. \ref{fig_scf}, left), the SCF is displayed along the void boundary for
the values of parameter $\Lambda=\left\{0.3,\sqrt{2}-1,0.8\right\}$ disclosing the number of stress annihilation points, which is respectively four, two, and zero.
In the other case, $\Upsilon=0.5$ (Fig. \ref{fig_scf}, right), the SCF is displayed for $\Lambda=0.75$ for three values of the angle difference $(\alpha-\beta)=\left\{1/4,3/4,1\right\}\pi$,
showing the presence
of stress annihilation at two points for the two lowest values and at no point for the highest value of the analyzed angle differences.

\subsection{Stress singularity and its removal at the isotoxal tips and the hypocycloid cusps}

\begin{figure}[!htb]
  \begin{center}
\includegraphics[width=16.8 cm]{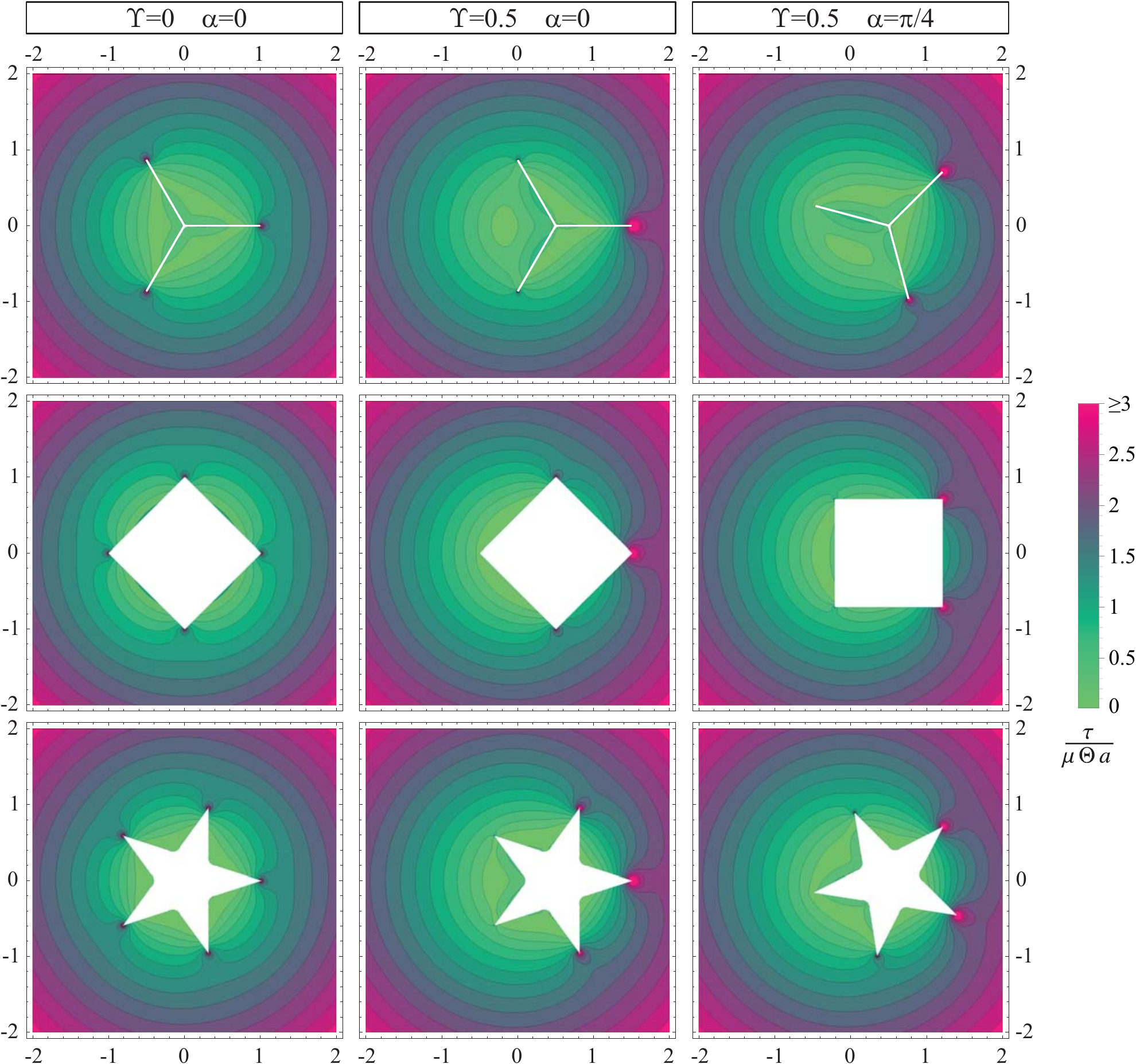}
\caption{Shear stress modulus $\tau$  (normalized through division by $\mu \Theta a$) around a three-pointed star-shaped crack (first line), a square (second line),
and a five-pointed regular star-polygon (third line). Different columns consider different void location, corresponding to $\Upsilon=\alpha=0$ (left column),
to $\Upsilon=0.5$ and $\alpha=0$ (central column), and to $\Upsilon=0.5$ and $\alpha=\pi/4$ (right column), with $\beta=0$ for all the cases.
\small
}
\label{fullfield_singular}
 \end{center}
\end{figure}

Stress singularities around cusps and points of sparsely distributed hypocycloidal and isotoxal voids are theoretically predicted from the full-field solutions obtained in the previous Section.
The strong stress intensification at the void tips is evident from Fig. \ref{fullfield_singular},
 where the shear stress modulus (normalized through division by $\mu \Theta a$)
is displayed  for different void geometry (shape and number $n$) and void location.
In particular, assuming $\beta=0$, three-pointed star-shaped cracks, squares, and five-pointed regular star-polygons are reported in the first, second, and third line of the figure, respectively, where
the case of void centroid coincident with the torsion axis ($\Upsilon=0$) is considered on the left column and the cases with $\Upsilon=0.35$ for $\alpha=0$ and $\alpha=\pi/4$ 
on the central and right column, respectively.

From Fig. \ref{fullfield_singular} it is noted that, while  polar symmetric stress distributions are displayed in the former case ($\Upsilon=0$) because of the polar symmetry in both geometry and loading,
the cases with $\Upsilon\neq0$ display different intensity
in the singularity at the different tips. To better elucidate this point, it is fundamental to analyze
the expressions of the Stress Intensity Factor  (SIF) and the Notch Stress Intensity Factor (NSIF) evaluated for
hypocycloidal voids, eqn (\ref{eq_sif_hypo}), and for isotoxal star-shaped polygonal voids, eqn (\ref{Kisotoxal}), (including the reduced cases of
$n$-sided regular polygonal void, eqn (\ref{NSIFpolygon}),
$n$-pointed regular star polygonal void, eqn (\ref{NSIFstar}),
and of $n$-pointed star-shaped cracks, eqn (\ref{eq_sif_star_crack})).
For all these cases, the SIF or NSIF  at the $k$-th cusp/tip can be represented by the following general expression
\beq\label{KKKgen}
K_{\text{III}}^{\left[\cdot\right]}(k) = \sqrt{\pi} \mu \Theta  a \mathcal{A}^{\left[\cdot\right]}(n) \left[\mathcal{C}^{\left[\cdot\right]}(n) +2 \Upsilon \cos \left(\alpha-\beta+\frac{2\pi(k-1)}{n}\right)\right],
\eeq
highlighting the superposition of the \lq pure' torsion (with axis of torsion corresponding to the centroid axis) and the uniform Mode III (Fig. \ref{fig2}), so that
\beq\label{ktrovati}
K_{\text{III}}^{\left[\cdot\right](t)} = \sqrt{\pi} \mu \Theta  a \mathcal{A}^{\left[\cdot\right]}(n) \mathcal{C}^{\left[\cdot\right]}(n),\qquad
K_{\text{III}}^{\left[\cdot\right](u)}(k) = 2\sqrt{\pi} \mu \Theta  a \Upsilon \mathcal{A}^{\left[\cdot\right]}(n) \cos \left(\alpha-\beta+\frac{2\pi(k-1)}{n}\right).
\eeq
The introduced parameters $\mathcal{A}^{\left[\cdot\right]}(n)$ and $\mathcal{C}^{\left[\cdot\right]}(n)$ are positive quantities depending
 on the void shape and on the number $n$ of vertexes/cusps, $\mathcal{A}^{\left[\cdot\right]}(n)$ is non-dimensionless and given by
\beq
\begin{array}{ccc}
\ds \mathcal{A}^{\left[\text{\ding{71}}\right]}(n)=\frac{\sqrt{ (n-1) a}}{n}, \qquad
\mathcal{A}^{\left[\text{\ding{85}}\right]}(n)=
2^{\frac{n(1-\xi)-2}{2n(1-\xi)}}
\left(\frac{ a \Omega^{\left[\text{\ding{85}}\right]}}{n (1-\xi)}\right)^{\frac{1-2\xi}{2(1-\xi)}},
\qquad
\mathcal{A}^{\left[\diamond\right]}(n)=
\sqrt{2}\left(\frac{ a \Omega^{\left[\diamond\right]}}{n +2}\right)^{\frac{2}{n+2}},
\\[5mm]
\ds\mathcal{A}^{\left[\text{\ding{73}}\right]}(n)=2^{\frac{n+8}{2(n+4)}}
\left(\frac{ a \Omega^{\left[\text{\ding{73}}\right]}}{n+4}\right)^{\frac{4}{n+4}},
\qquad
\mathcal{A}^{\left[\ast\right]}(n)=2 ^{\frac{n-4}{2n}}\sqrt{\frac{ a}{n}},
\end{array}
\eeq
while $\mathcal{C}^{\left[\cdot\right]}(n)$ is dimensionless and given by
\beq\label{matcalCC}
\begin{array}{ccc}
\ds \mathcal{C}^{\left[\text{\ding{71}}\right]}(n)=1, \qquad
\mathcal{C}^{\left[\text{\ding{85}}\right]}(n)=
n   \Omega^{\left[\text{\ding{85}}\right]}\ds
\ds\sum_{j=0}^{\infty}\sum_{l=0}^{j} (j-l) d_j^{\left[\text{\ding{85}}\right]} d_l^{\left[\text{\ding{85}}\right]},
\\[8mm]
\mathcal{C}^{\left[\diamond\right]}(n)=
n   \Omega^{\left[\diamond\right]}\ds
\ds\sum_{j=0}^{\infty}\sum_{l=0}^{j} (j-l) d_j^{\left[\diamond\right]} d_l^{\left[\diamond\right]},
\qquad
\mathcal{C}^{\left[\text{\ding{73}}\right]}(n)=
n   \Omega^{\left[\text{\ding{73}}\right]}\ds
\ds\sum_{j=0}^{\infty}\sum_{l=0}^{j} (j-l) d_j^{\left[\text{\ding{73}}\right]} d_l^{\left[\text{\ding{73}}\right]},
\\[16mm]
\mathcal{C}^{\left[\ast\right]}(n)=\ds\dfrac{ 2}{n} \left[\Gamma\left(\frac{2}{n}\right)\right]^2\ds
\ds\sum_{j=0}^{\infty}
\frac{ {}_2F_1\left[j,1+\frac{2}{n};2+j;-1\right]}{\Gamma\left(1+j\right)\Gamma\left(2+j\right)\Gamma\left(\frac{2}{n}-j\right)\Gamma\left(1-j+\frac{2}{n}\right)}.
\end{array}
\eeq

Values of the parameter $\mathcal{C}^{\left[\cdot\right]}(n)$ assessed for
$n$-sided regular polygonal voids, for $n$-pointed regular star polygonal voids, and  for
$n$-pointed star-shaped cracks are reported in Tab. 1 for values of $n$ up to ten.\footnote{The evaluation of   $\mathcal{C}^{\left[\cdot\right]}(n)$, and therefore that of
SIF and NSIF, in the case of isotoxal star-shaped polygonal voids requires in general the  computation of a series, of which convergence is discussed in Appendix A.
}

\begin{table} [!ht]
\label{cenne}
\begin{center}
\begin{tabular}{c}
\toprule[.8pt]
$\mathcal{C}^{\left[\cdot\right]}(n)$\\
\begin{tabular}{c|cccc}
\toprule[.8pt]
$n$  & $n$-sided &  $n$-pointed & $n$-pointed \\[1mm]
 & regular polygon $\left[\diamond\right]$ &  regular star polygon  $\left[\text{\ding{73}}\right]$ & star-shaped crack $\left[\ast\right]$\\
\hline
  $3$        &    1.06353      &     -        & 0.92037    \\[1mm]
  $4$        &    1.04484      &     -        & 0.90032    \\[1mm]
  $5$        &    1.00924      &     0.97293  & 0.89663    \\[1mm]
  $6$        &    0.96811      &     1.00270  & 0.89854    \\[1mm]
  $7$        &    0.92550      &     1.01021  & 0.90249    \\[1mm]
  $8$        &    0.88345      &     1.00726  & 0.90708    \\[1mm]
  $9$        &    0.84302      &     0.99843  & 0.91174    \\[1mm]
  $10$       &    0.80473      &     0.98594  & 0.91623    \\[1mm]
\bottomrule
\end{tabular}
\end{tabular}
\bf \small \caption{\textnormal{Values of $\mathcal{C}^{\left[\cdot\right]}(n)$ assessed for different void shape and number of tips  $n$, fundamental to tailor the radial distance parameter $\Upsilon$
to achieve stress singularity removal at some cusp as defined by eqn (\ref{removal}).
} }
\end{center}
\end{table}

The general expression (\ref{KKKgen}) mathematically shows that only when the radial distance is null ($\Upsilon=0$, value providing the mentioned polar symmetric condition)
the singularity at each cusp/point of the void has the same intensity,
\beq
K_{\text{III}}^{\left[\cdot\right]}(k) =
K_{\text{III}}^{\left[\cdot\right](t)}\qquad
\forall\, k\in[1,n] \qquad
\iff \qquad
\Upsilon=0.
\eeq
and on the other hand, the obtained expressions for $K_{\text{III}}^{\left[\cdot\right](u)}(k)$, eqn (\ref{ktrovati})$_2$, independently confirm
the SIF evaluated in \cite{partII} when the polynomial loading condition
is restricted to Uniform Mode III.

Equation  (\ref{KKKgen}) also discloses that
the stress singularity at each cusp/point of the void is characterized by a different intensity
in the presence of a non-null radial distance ($\Upsilon \neq 0$).
In this case, the void position and inclination can be
tailored through the parameters of radial distance ratio $\Upsilon$ and angular difference $\alpha-\beta$
  towards the singularity decrease, or even, removal at some cusp/point of the void, namely, $K_{\text{III}}(k)=0$ for specific $k$.
Three cases can be therefore distinguished:\footnote{Without loss of generality, when possible, the singularity removal is considered to occur
at the point corresponding to $k=1$. Having restricted the analysis to regular symmetric void geometries, the point numbering has no special meaning except
in defining  the point position with respect to the $\widehat{x}_1$ axis.}
\begin{itemize}
\item Case $\Upsilon<\mathcal{C}^{\left[\cdot\right]}(n) /2$. All the cusps/points are characterized by a SIF varying in its magnitude but not in its sign. No removal of singularity is possible;
\item Case $\Upsilon=\mathcal{C}^{\left[\cdot\right]}(n) /2$. Singularity disappears at the cusp/point $k$=1, for the special value of angular difference
\beq
\alpha-\beta=(2j+1)\pi,\,\,\,\, j\in \mathbb{Z} \qquad  \Rightarrow \qquad K_{\text{III}}(k=1)=0.
\eeq
Differently, if $\alpha-\beta\neq(2j+1)\pi$ (with $j\in \mathbb{Z}$), the SIFs or NSIFs at all the cusps/points have the same sign.

\item Case $\Upsilon>\mathcal{C}^{\left[\cdot\right]}(n) /2$. In this case, the cusps/points can be collected in two sets depending on the sign of their SIF/NSIF. Singularity removal may occur
at one cusp/point or at two cusps/points.
\begin{itemize}
\item Singularity is removed at the two cusps/points corresponding to $k=1$ and $k=1+m$, when both the following conditions hold
\beq\label{removal}
\left\{
\begin{array}{lll}
\ds\alpha-\beta=\pi\left(1+2j-\frac{m}{n}\right),\\[3mm]
\Upsilon=\ds\frac{\mathcal{C}^{\left[\cdot\right]}(n) }{\ds 2\cos\frac{m \pi}{n}}
\end{array}
\right.
\Rightarrow
\left\{
\begin{array}{lll}
K_{\text{III}}(k=1)=0,\\[3mm]
K_{\text{III}}(k=1+m)=0,
\end{array}
\right.
\,\,\,\,\,\,\,\,
m=1, ..., \left\lfloor \frac{n-1}{2}\right\rfloor,\,\,\,\,
j\in \mathbb{Z},
\eeq
where the symbol $\lfloor \cdot \rfloor$ provides the integer part of the relevant argument.
\item  Differently, if  $\alpha-\beta\neq\pi[1+2j-m/n]$, the singularity disappears only at the cusp/point corresponding to $k$=1 for  the following angular difference
\beq
\alpha-\beta=\arccos\left[-\frac{\mathcal{C}^{\left[\cdot\right]}(n) }{2\Upsilon}\right]\neq \pi\left[1+2j-\frac{m}{n}\right]
\Rightarrow  K_{\text{III}}(k=1)=0,
\,\,\,\,\,\,\,\,
m=1, ..., n,
\,\,\,\,\,\,\,\,
j\in \mathbb{Z},
\eeq
\item Otherwise, no singularity removal occurs at any cusp/point.
\end{itemize}
\end{itemize}

Note that equation (\ref{removal}) provides  a number $\left\lfloor (n-1)/2\right\rfloor$
of independent conditions for which different pairs of vertexes ($k=1$ and $k=1+m$) display simultaneously  singularity removal.
Such independent conditions have been obtained restricting the parameter $m$ to the set $[1, \left\lfloor (n-1)/2\right\rfloor]$,
because values of $m$ within the set $[\left\lfloor (n-1)/2\right\rfloor +1, n]$  provide singularity removal conditions for a
negative distance parameter $\Upsilon$ and merely correspond to a reflection of the configurations associated to the independent conditions.

\begin{figure}[!h]
  \begin{center}
\includegraphics[width=16.8 cm]{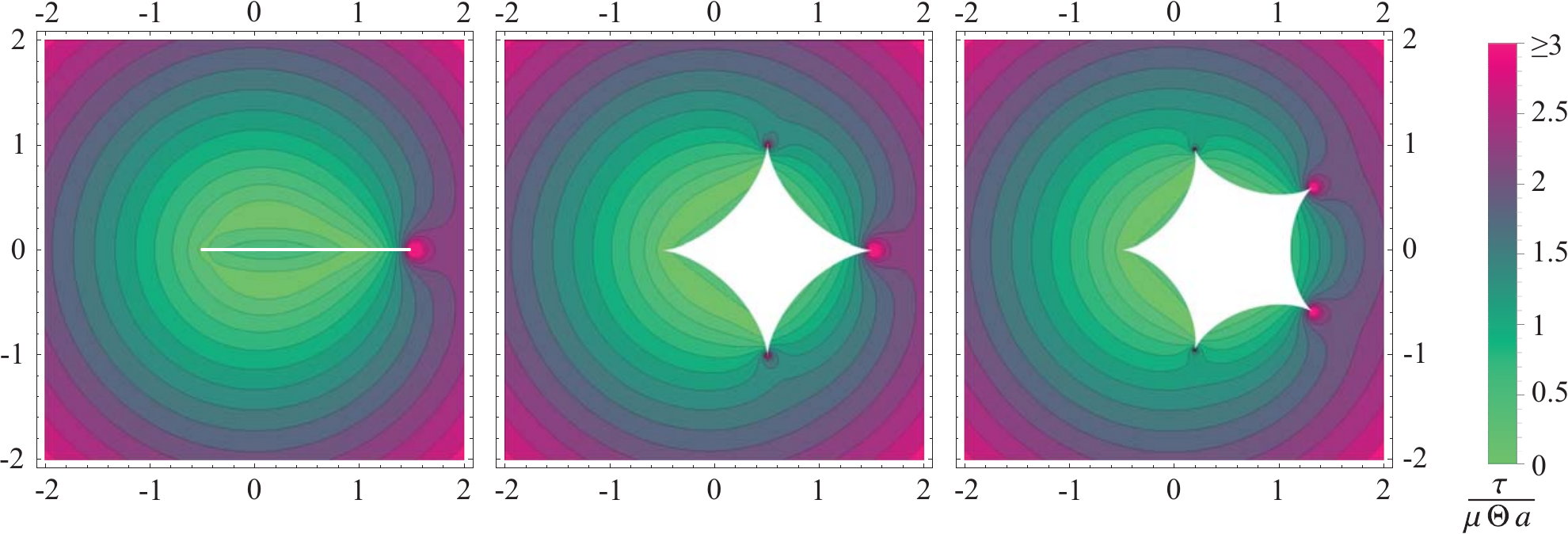}
\caption{Stress singularity removal at one tip of a crack (left) and at one cusp of a four- (center) and five- (right)
cusped hypocycloidal shaped inclusion attained for a dimensionless radial distance $\Upsilon=0.5$ and $\alpha=\pi$ (with $\beta=0$).
The contours of shear stress modulus $\tau$ are reported normalized through division by $\mu\Theta a$.
\small
}
\label{fullfield_1}
 \end{center}
\end{figure}

\begin{figure}[!h]
  \begin{center}
\includegraphics[width=16.8 cm]{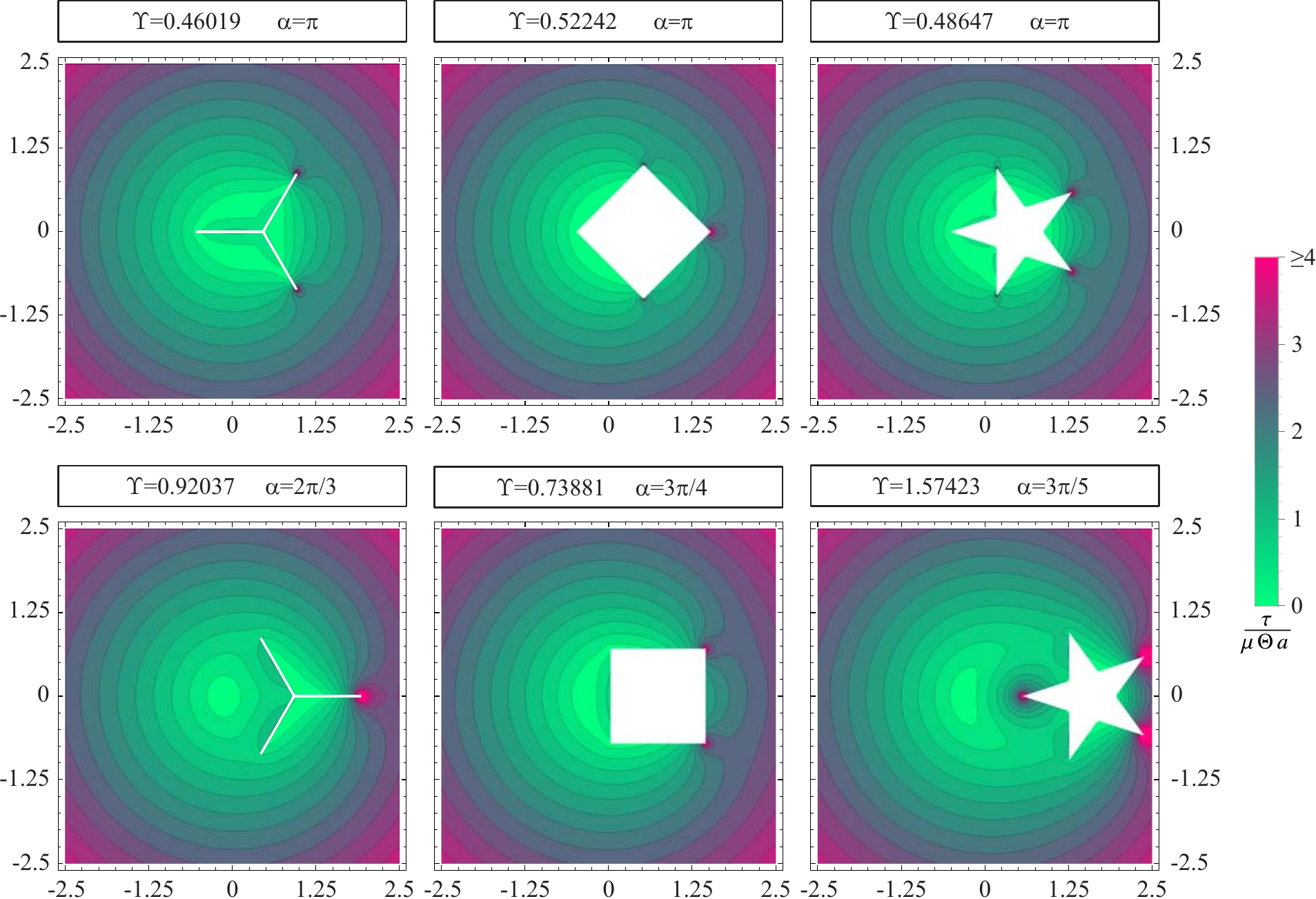}
\caption{Stress singularity removal at one point (first line) and at two points (second line) of the same isotoxal void. A three-pointed crack (left), a square (center)
and a five-pointed star (right) are considered. The special values for which this behaviour is attained for each void shape are reported  in the figure (where $\beta=0$ is assumed).
The contours of shear stress modulus $\tau$  are reported normalized through division by $\mu\Theta a$.
\small
}
\label{fullfield_2}
 \end{center}
\end{figure}

Examples of singularity removal at one and two cusps/points are reported in Fig. \ref{fullfield_1} and in \ref{fullfield_2}.
More specifically, a radial normalized distance $\Upsilon=0.5$ (and angles $\alpha=\pi$ and $\beta=0$) is considered in  Fig. \ref{fullfield_1} to achieve stress singularity removal at one tip of a \lq standard' crack and at one cusp of a four and a five-cusped
hypocycloidal shaped inclusion. The three void shapes share the same centroid position for achieving the singularity removal, being the dimensionless parameter the same for the \lq standard' crack and for any $n$-cusped hypocycloidal shaped inclusion,
$\mathcal{C}^{\left[\text{\ding{71}}\right]}(n)=\mathcal{C}^{\left[-\right]}=1$.
In  Fig. \ref{fullfield_2} the singularity removal is attained  at one point (first line) and two points (second line) for the same inclusion shape, a three-pointed crack (left column),
a square (central column), and a five-pointed star (right column). This feature is observed for the specific values of $\Upsilon$ and $\alpha$ listed in the figure (keeping $\beta=0$).


It is worth remarking that, for the point where stress singularity removal occurs, the leading order term in the stress asymptotic representation becomes a positive power of the vanishing radial distance from the point of an isotoxal shar-shaped polygon, while it becomes a constant (called T-stress under in-plane conditions) for an hypocycloidal shaped void and for a star-shaped crack. Therefore, as observed in \cite{partI}, the singularity removal at the point of an isotoxal shar-shaped polygon void also implies the stress annihilation, while singularity removal at a cusp or at a tip does not necessarily imply a null stress state (because of the constant stress term in the related asymptotic expansion). Nevertheless,  a null stress state is numerically verified also in this latter case (while such analytical proof seems awkward).

Finally, although the theoretical sets of the distance $\Upsilon$ and angular difference $\alpha-\beta$ leading to singularity removal
expressed by eqn (\ref{removal}) have been disclosed in the limit case of an infinite elastic matrix, the obtained results
are also very indicative in practical cases where cross sections are defined by finite regions, as numerically shown in the next Section.

\section{Accuracy assessment of the analytical expressions in finite domain applications}

Finite Element simulations are performed towards the accuracy assessment in using the presented theoretical predictions (derived for a single void in an infinite medium  in Sect \ref{ellipsesection})
for practical realizations, where the cross section has finite dimensions.
Within a planar setting, the numerical results are obtained  for the case of doubly connected uniform cross section where the influence of cross section shape and size is
evaluated.
The stress function $\psi$ for a doubly connected domain with an external boundary $\mathcal{B}_0$ (defining the external shape of the cross section)
 and an internal boundary $\mathcal{B}_1$ (defining the
void geometry) is given by \cite{chenyz2}
        \beq\label{torcomsol}
        \psi(x_1,x_2) =\psi_0(x_1,x_2)\,\, -\,\,\ds \frac{\ds\oint_{\mathcal{L}} \frac{\partial \psi_0(x_1,x_2)}{\partial n} \mathrm{d}s}{\ds\oint_{\mathcal{L}} \frac{\partial \psi_1(x_1,x_2)}{\partial n} \mathrm{d}s} \,\,\psi_1(x_1,x_2),
        \eeq
where
the functions $\psi_0$ and $\psi_1$, in addition to the Laplacian equation ($\nabla^2 \psi_0=\nabla^2 \psi_1=0$), are subject to the following boundary conditions
\beq
\ds\left.\psi_0(x_1,x_2)\right|_{\mathcal{B}_0}=\frac{x_1^2+x_2^2}{2},\quad
\left.\psi_0(x_1,x_2)\right|_{\mathcal{B}_1}=\frac{x_1^2+x_2^2}{2},\qquad
\ds\left.\psi_1(x_1,x_2)\right|_{\mathcal{B}_0}=0,\quad
\left.\psi_1(x_1,x_2)\right|_{\mathcal{B}_1}=1,
\eeq
$\mathcal{L}$ is any (closed) contour between boundary $\mathcal{B}_0$ and $\mathcal{B}_1$, and the directional derivative of the
function $\psi$ along the outward normal vector $n=\left\{n_1,n_2\right\}$ orthogonal to the contour $\mathcal{L}$, can be expressed as
\beq
\frac{\partial \psi}{\partial n}=\frac{\partial \psi}{\partial x_1} n_1+\frac{\partial \psi}{\partial x_2}n_2.
\eeq

Restricting for simplicity to null angle values, $\alpha=\beta=0$, the above formulation for the torsion problem is  numerically solved through the \lq Equation-based Modeling' feature of the
 Mathematics module in Comsol Multiphysics$^\copyright$ version 5.3. The stress analysis is performed under stationary conditions in the presence of an elliptical void, a star-shaped crack or an hypocycloidal hole
 (enclosed by the smallest circle of radius $a$)
 in an elastic matrix with different shape and size, the latter defined through the length $D$.
The whole domain is meshed using the user-controlled mesh option with custom free triangular element size at two levels.
At first level, the domain is meshed with maximum element size of $0.01 D$, and at the second level, refined mesh is used for the voids boundary with
maximum element size of $0.0025 D$.
Accuracy in the evaluation of intensification factors is checked through the convergence of the relevant quantities, which is considered reached when a difference less than 0.1$\%$
is evaluated between two successive automatic refinements of the mesh.

\subsection{Elliptical voids}

Two types of comparison are reported at varying ellipse geometry and external boundary size and shape, with the ellipse centroid  coincident to the torsion axis ($\Upsilon=0$).

\begin{figure}[!htp]
  \begin{center}
\includegraphics[width=16 cm]{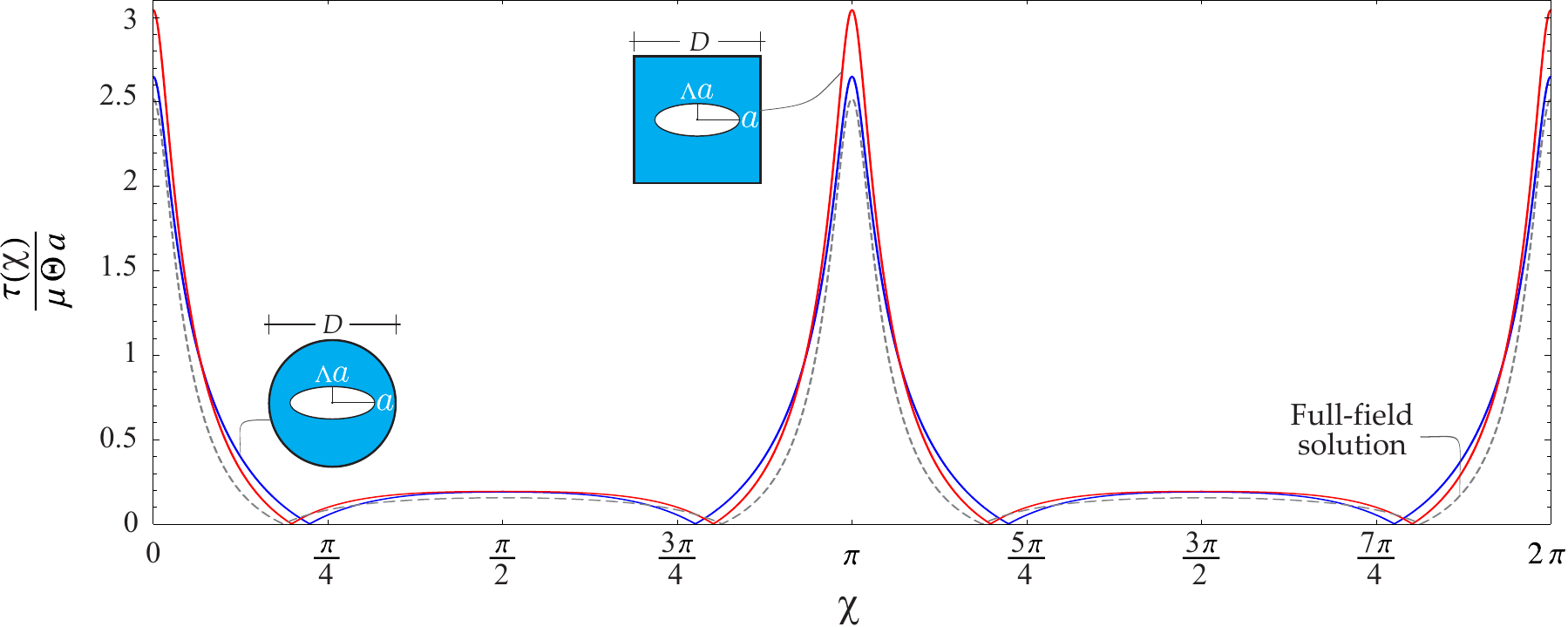}
\caption{
\small
Shear stress modulus $\tau$ (normalized through division by $\mu \Theta a$)  along the  void boundary for elliptical hole with aspect ratio $\Lambda=0.3$
evaluated through the presented full-field solution (dashed line) and the  finite element simulations (continuous lines) performed in the cases of a  circular and a square elastic matrix with a small size, $D/(2a)=1.5$.
The stress is reported as a function of the polar counter-clockwise angle $\chi$, with $\chi=0$ corresponding to the ellipse point $x_1=a$, $x_2=0$.
Despite the considered smallness in the domain size, the numerical simulations display the presence along the void boundary of four stress annihilation points  very close to those predicted by the full-field solution. 
}
\label{SCF_ellipse2}
 \end{center}
\end{figure}

The first comparison is about the shear stress modulus $\tau(\chi)$  along the boundary of an elliptical void with $\Lambda=0.3$ in a finite domain with size  $D/(2a)=1.5$,  Fig. \ref{SCF_ellipse2}.
The  prediction from the presented full-field solution about the presence of four stress annihilation points  is confirmed by
the finite element simulations in both the cases of circular and square domain. In particular, despite the small size of the considered finite domains,
 only a small change is observed in the value of the  polar angle $\widetilde{\chi}$ where a null stress is attained, $\tau(\widetilde{\chi})=0$.

\begin{figure}[!h]
  \begin{center}
\includegraphics[width=17 cm]{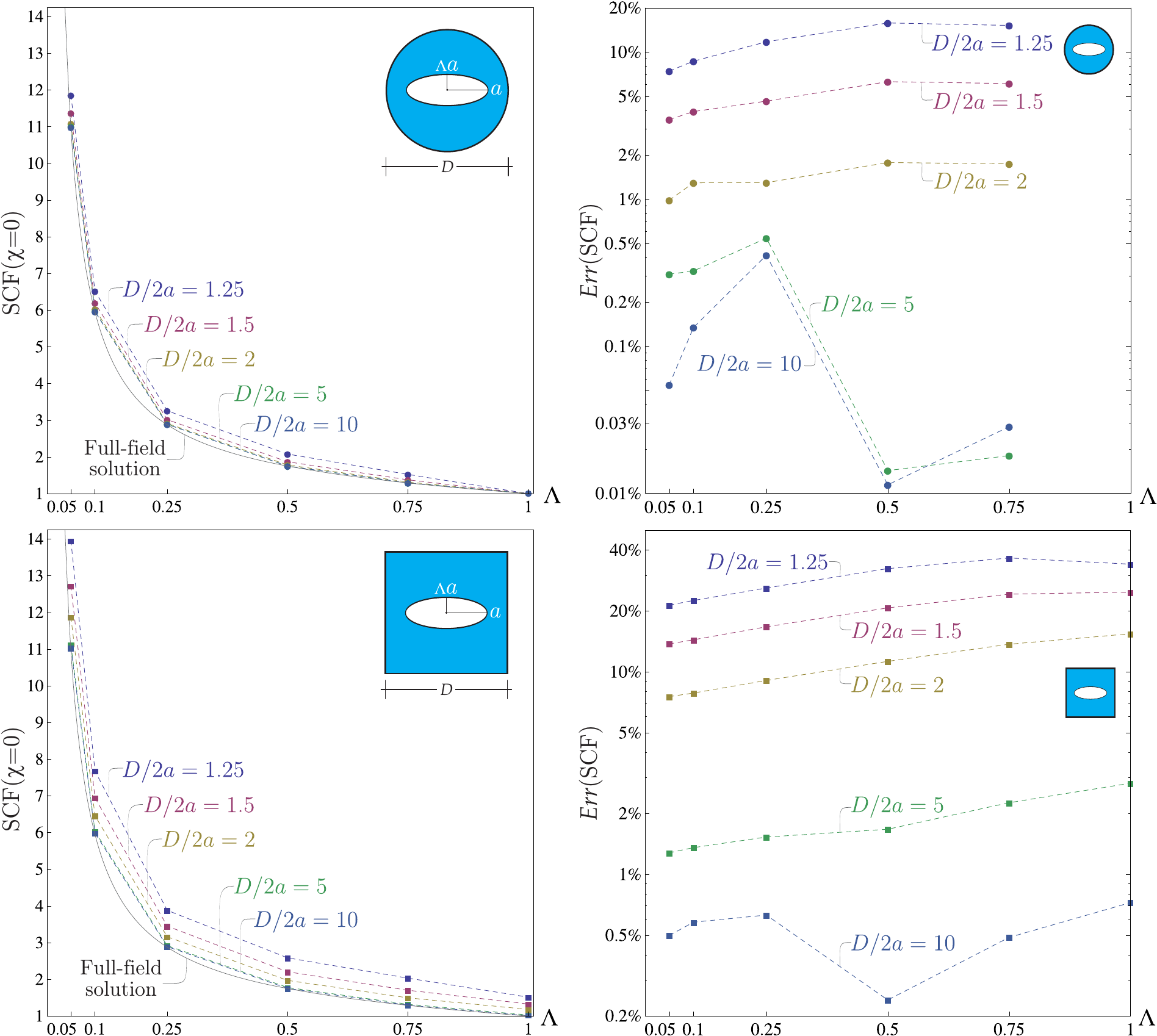}
\caption{
\small SCF evaluated at the major axis ($\chi=0$) of an elliptical void in a circular (upper part) and a square (lower part) elastic domain.
(Left column) Comparison between the analytical expression (continuous curve), eqn (\ref{reducedSCF}), and the numerical values (dots)
obtained from finite element simulations performed for different ellipse parameter $\Lambda=\left\{0.05, 0.1, 0.25, 0.5, 0.75, 1 \right\}$ and different matrix size,
$D/(2a)=\left\{1.5,2,5,10\right\}$. (Right) For the same geometry conditions considered in the left column,
the error (reported in logarithmic scale)
made by assuming the analytical expression obtained for the void in an infinite elastic plane assessed through the quantity $Err(\text{SCF})=|\text{SCF}_{FE}-\text{SCF}|/\text{SCF}_{FE}$.
}
\label{SCF_ellipse}
 \end{center}
\end{figure}

The second comparison is  displayed in Fig. \ref{SCF_ellipse} and is based on the Stress Concentration Factor (SCF) attained at
the ellipse major axis, which is theoretically given, by  reducing eqn (\ref{scfchizero}) with the parameter $\chi=0$, as
\beq\label{reducedSCF}
\text{SCF}(\chi=0)=\frac{1}{2}  \left(2+\frac{1}{\Lambda } -\Lambda\right).
\eeq
The analytical predictions for the SCF are compared to those numerically evaluated $\text{SCF}_{FE}$ from the finite element simulations at different ellipse parameter $\Lambda=\left\{0.05,0.1,0.25,0.5,0.75,1\right\}$
 for a circular (Fig. \ref{SCF_ellipse}, upper part) and square (Fig. \ref{SCF_ellipse}, lower part)  boundary of the elastic matrix with different size, $D/(2a)=\left\{1.5,2,5,10\right\}$.
The SCF values are displayed on the left column, while the error made by assuming the analytical prediction (and defined as $Err(\text{SCF})=|\text{SCF}_{FE}-\text{SCF}|/\text{SCF}_{FE}$) is reported (in logarithmic scale) on the right column.
In the case $\Lambda=1$ and circular domain (upper part, right), corresponding to the
special geometry of annular cross section, the full-field solution for the infinite domain coincides with that for the finite domain so that $Err(\text{SCF})$ is null under this condition (and not displayed because outside the plotrange).
From the right column of Fig. \ref{SCF_ellipse}, it can be observed that $Err(\text{SCF})$ increases at decreasing the size $D$, however the error using the analytical expression (\ref{reducedSCF}) is
below $2\%$ for circular domains with $D/(2a)\geq 2$
and below $3\%$ for square domains with $D/(2a)\geq 5$.

\subsection{Star-shaped cracks and hypocycloidal holes}

\subsubsection{$J$-integral extension to the torsion problem}

The  $J$-integral \cite{r} is a conservative quantity thoroughly exploited over the years in  crack problems under in-plane and out-of-plane  conditions towards the evaluation of the Stress Intensity Factor (SIF)
for  specific boundary value problems. However,
for the problem under consideration, the classical definition of $J$-integral can not be applied because torsion loading conditions do not realize neither \lq pure' in-plane nor \lq pure' out-of-plane states.

Towards the definition of a conservative integral for the considered problem,
reference is made to the conservative integral for three-dimensional linear elastic solids introduced by Knowles and Sternberg \cite{j} (their eqn 3.16) which,
 taking into account of the vanishing kinematical and stress quantities in the torsion problem, reduces to
\beq
\label{eq_j_torsion}
J_{torsion}=\ointctrclockwise_{\mathcal{L}} \left[\frac{\widehat{\tau}_{13}^2+\widehat{\tau}_{23}^2}{2 \mu} \widehat{n}_1 -
\left(\widehat{\tau}_{13} \widehat{n}_1+\widehat{\tau}_{23} \widehat{n}_2\right)\frac{\partial  \widehat{v}_3(\widehat{x}_1,\widehat{x}_2)}{\partial \widehat{x}_1}\right]\mathrm{d} s
-\int_{\mathcal{S}} \widehat{\tau}_{23}(\widehat{x}_1,\widehat{x}_2)\mathrm{d} \mathcal{S}.
\eeq
In equation (\ref{eq_j_torsion}), $\mathcal{L}$ is any counterclockwise contour enclosing the crack tip lying along the $\widehat{x}_1$ axis, $s$ is the curvilinear coordinate along the
contour $\mathcal{L}$, $\mathcal{S}$ is the region enclosed by the contour $\mathcal{L}$, $\widehat{n}_1$ and $\widehat{n}_2$ denote the Cartesian components of the outward unit normal
to the contour $\mathcal{L}$ along the $\widehat{x}_1$ and $\widehat{x}_2$ directions. Therefore, it can be noted that the conservative integral for the torsion problem differs from that for Mode III loading conditions,
$J_{torsion}\neq J_{III}$. More precisely, a surface integral is present in equation (\ref{eq_j_torsion}) in addition to the contour integral, which is coincident with the definition of $J$-integral under Mode III (antiplane, or out-of-plane) loading conditions, $J_{III}$.

Considering the asymptotic behaviour of the kinematical and stress fields, equation (\ref{eq_j_torsion}) reduces to the following relation connecting the conservative  integral to the SIF
\beq\label{JvsK}
J_{torsion}=\frac{K_{\text{III}}^2}{2 \mu}.
\eeq
Equation (\ref{JvsK}) represents a key tool in the  evaluation of the SIF and, used
in combination with results from finite element simulations, allows for assessing the accuracy in using the analytical expressions for practical realizations.

\subsubsection{Star-shaped cracks and hypocycloidal shaped  voids in bounded domains}

Stress Intensity Factors $K_{\text{III}}$ have been numerically evaluated for different void geometry and different cross section shape and size, defined by the length $D$.
Using eqn (\ref{JvsK}),  $K_{\text{III,FE}}$ have been obtained through the numerical evaluation of  the  $J$-integral (\ref{eq_j_torsion}) computed from finite element simulations
 using a square contour enclosing the tip of the void.\footnote{The size and the center of the square used for the computation of the $J$-integral have been considered different for
 the different geometries of the void. In particular, the square considered for star-shaped cracks  has the side equal to  $a/2$ and center located at the crack tip, while
 for hypocycloidal shaped void has the side equal to $a/8$ and centered along the symmetry axis of the relevant cusp at a distance equal to $a/23$
 (in order to limit the error in computing the conservative integral (\ref{eq_j_torsion}) generated by the presence of a non-null curvature in the hypocycloidal shaped  void at the cusp).
 } The numerical evaluations are compared with the
analytical values $K_{\text{III}}$ obtained under the assumption of
infinite elastic matrix in order to assess the reliability of the
presented results for applicative problems. Four main
problems are considered and discussed.

\begin{figure}[!htb]
  \begin{center}
\includegraphics[width=18 cm]{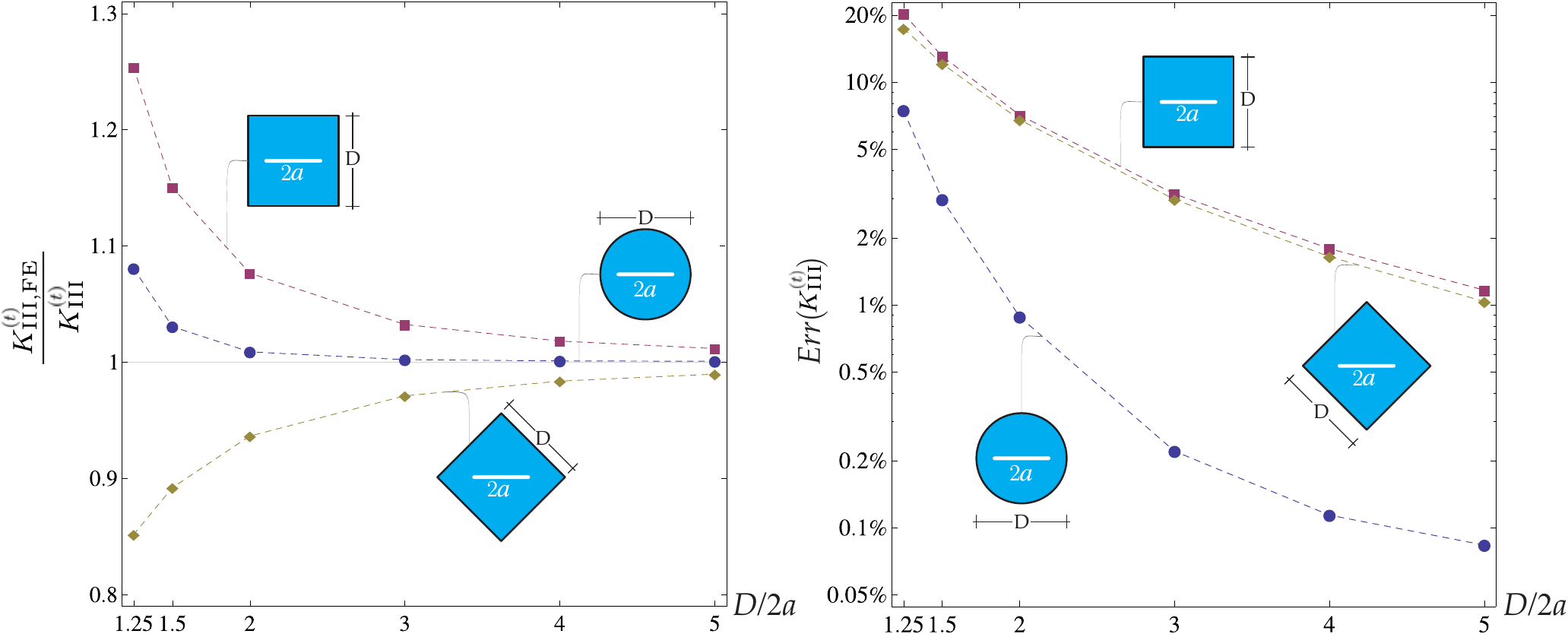}
\caption{
\small (Left) Numerical evaluation of Stress Intensity Factor $K_{\text{III,FE}}^{(t)}$ (normalized through
division by the corresponding analytical value obtained for the infinite elastic plane, $K_{\text{III}}^{(t)}=\sqrt{\pi}\mu \Theta a^{3/2}/2$) at varying the size and shape of the elastic matrix domain. (Right)
For the same geometry conditions considered in the left column, the error (reported in logarithmic scale)
made by assuming the analytical expression obtained for the void in an infinite elastic plane assessed through the quantity $Err(K_{\text{III}}^{(t)})=|K_{\text{III,FE}}^{(t)}-K_{\text{III}}^{(t)}|/K_{\text{III,FE}}^{(t)}$.
}
\label{KR_Kinf}
 \end{center}
\end{figure}

\paragraph{Standard crack  with centroid coincident to the torsion axis, $\Upsilon=0$.} The Stress Intensity Factor $K_{\text{III,FE}}^{(t)}$
is numerically evaluated for different matrix boundary (circular, square with sides parallel and orthogonal to the crack line, and square with sides inclined at an angle $\pi/4$ with respect to
the crack line) and size, $D/(2a)=\left\{1.25,1.5,2,3,4,5\right\}$. This quantity, normalized through division by the corresponding analytical value
for the infinite matrix,  $K_{\text{III}}^{(t)}=\sqrt{\pi}\mu \Theta a^{3/2}/2$, is reported in Fig. \ref{KR_Kinf} (left), and used to assess
the error  $Err(K_{\text{III}}^{(t)})=|K_{\text{III,FE}}^{(t)}-K_{\text{III}}^{(t)}|/K_{\text{III,FE}}^{(t)}$ reported in Fig. \ref{KR_Kinf} (right) in logarithmic scale.
Similarly to the case of elliptical void, the value of $Err(K_{\text{III}}^{(t)})$ increases at decreasing the size $D$ and the error in using the analytical expression (\ref{reducedSCF}) is
below $1\%$ for circular domains with $D/(2a)\geq 2$
and below $3\%$ for square domains with $D/(2a)\geq 4$.
The different  ranges for the size in the two cases are due to the  additional warping originated by  the external boundary, much more present for non-smooth external boundaries than for smooth ones.

\begin{figure}[!htb]
  \begin{center}
\includegraphics[width=16 cm]{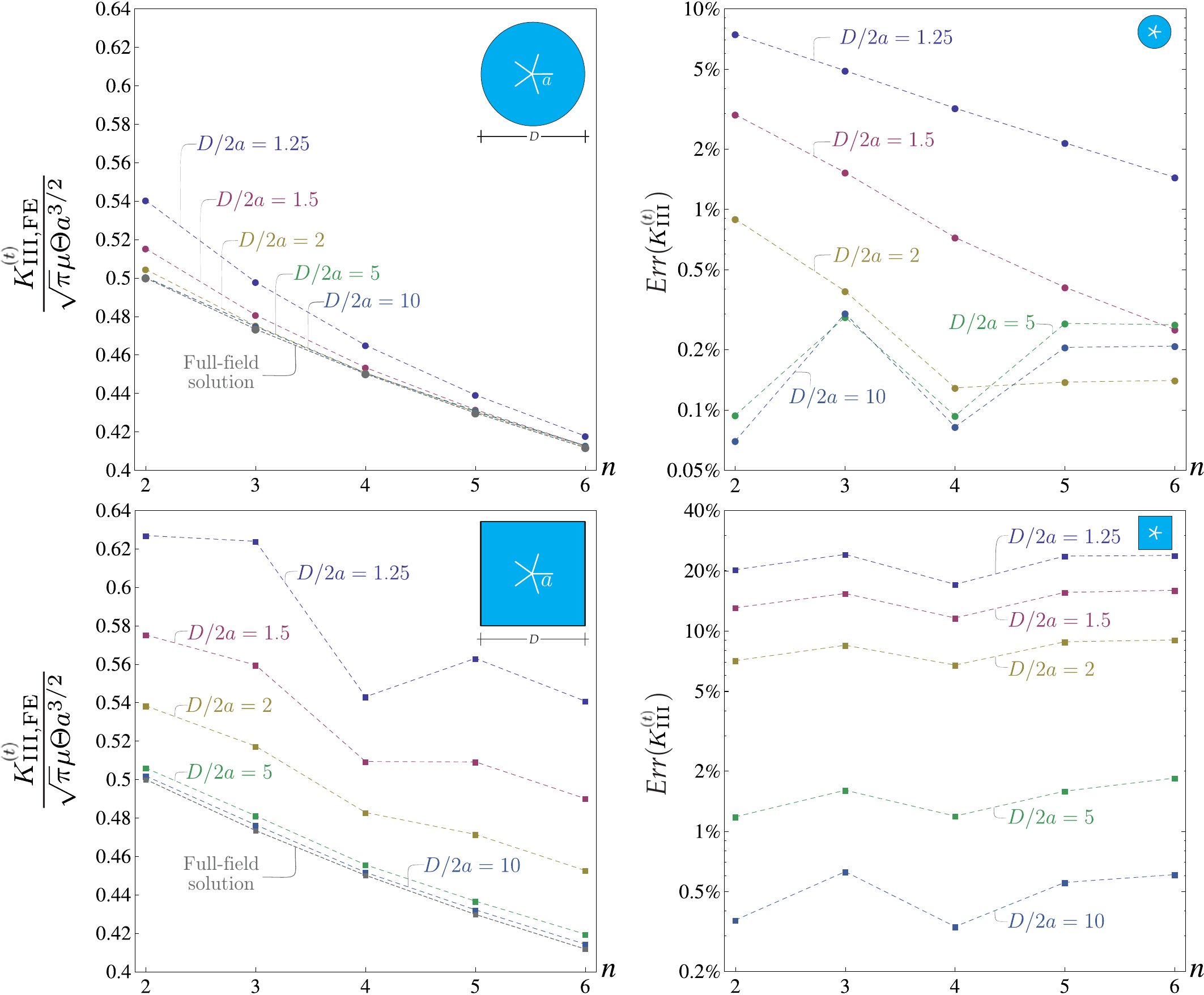}
\caption{
\small (Left) Comparison of the analytical and numerical values of the SIF (normalized through division by $\sqrt{\pi}\mu\Theta a^{3/2}$) at the tips of
$n$-pointed star-shaped cracks in a circular (upper part, left) and a square (lower part, left) elastic matrix subject to a \lq pure' torsion loading ($\Upsilon=0$).
(Right) For the same geometry conditions considered in the left column, the error (reported in logarithmic scale)
made by assuming the analytical expression obtained for the void in an infinite elastic plane assessed through the quantity $Err(K_{\text{III}}^{(t)})=|K_{\text{III,FE}}^{(t)}-K_{\text{III}}^{(t)}|/K_{\text{III,FE}}^{(t)}$.
}
\label{validation_1}
 \end{center}
\end{figure}

\paragraph{$n$-pointed star-shaped crack with centroid coincident to the torsion axis, $\Upsilon=0$.}
The Stress Intensity Factor $K_{\text{III, FE}}^{(t)}$ (normalized through division by $\sqrt{\pi}\mu\Theta a^{3/2}$)
is evaluated at varying the points number $n$ for a star-shaped crack in an elastic matrix with different size and boundary, circular (Fig. \ref{validation_1}, upper part, left)
and square with sides parallel and orthogonal to one of the crack lines
(Fig. \ref{validation_1}, lower part, left).
Similarly, to the previous case, the error in using the analytical expression
for $K_{\text{III}}^{(t)}$, eqn (\ref{eq_sif_star_crack}) with $\Upsilon=0$,
is displayed on the right column for the respective case through the quantity $Err(K_{\text{III}}^{(t)})=|K_{\text{III,FE}}^{(t)}-K_{\text{III}}^{(t)}|/K_{\text{III,FE}}^{(t)}$ (reported in logarithmic scale).
It can be observed that at varying the void geometry, as in the previous case, the error remains below $1\%$ for circular domains with $D/(2a)\geq 2$
and below $3\%$ for square domains with $D/(2a)\geq 4$.

\begin{figure}[!htb]
  \begin{center}
\includegraphics[width=18 cm]{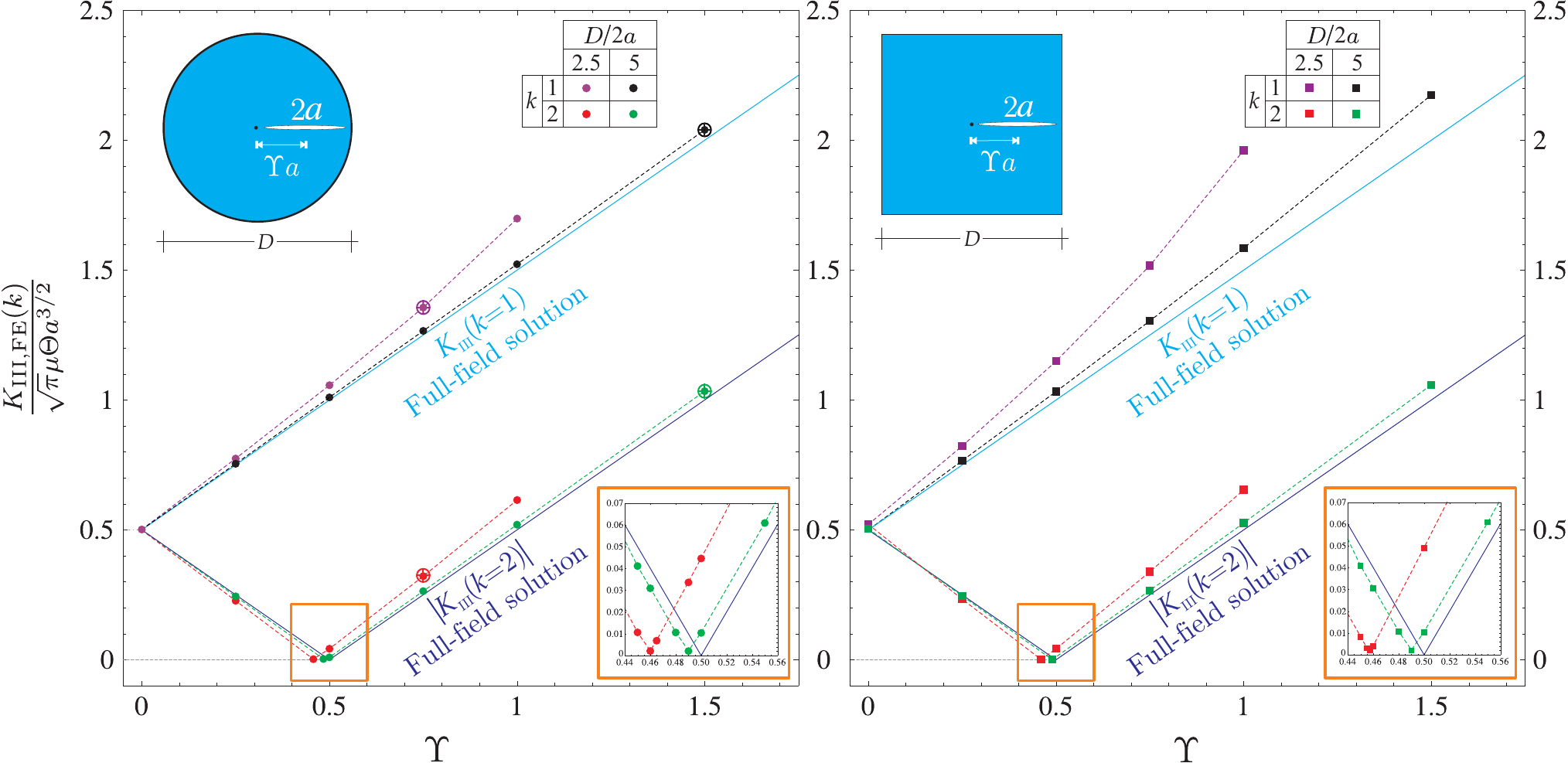}
\caption{
\small Comparison between analytical and numerical evaluation of the SIF at the two tips of a \lq standard' crack in an elastic domain at varying the distance parameter $\Upsilon$. Numerical evaluation is performed for different shape and size of the elastic matrix.
A strong reduction (about 215 times) in the modulus of  $K_{\text{III, FE}}(k=2)$ is observed for specific values of the distance parameter $\Upsilon \backsimeq0.5$ as the practical realization of the stress singularity
removal theoretically predicted from the presented full-field solution, as shown in Fig. \ref{fullfield_1} (left). The four empty circles shown on the left part for the two crack tips ($k=1$ and $k=2$) are referred to the corresponding values in Tab. 2, evaluated by \cite{iran2} for circular cross sections with $D/(2a)=10\Upsilon/3$, in the cases $\Upsilon=0.75$ (so that $D/(2a)=2.5$) and $\Upsilon=1.5$ (so that $D/(2a)=5$). 
}
\label{Knullity_crack_Circle}
 \end{center}
\end{figure}

\paragraph{A standard crack at varying the centroid position, $\Upsilon\neq0$.}
The SIFs at the two tips of a \lq standard' crack, numerically evaluated from the finite element simulations as $K_{\text{III, FE}}(k=1)$ and $K_{\text{III, FE}}(k=2)$,
are compared with the corresponding values from analytical expression (\ref{sif_crack_n2}) at varying the dimensionless radial distance
$\Upsilon$ in the case of circular and square elastic matrix in Fig. \ref{Knullity_crack_Circle}, left and right, respectively. Focussing attention to the singularity removal feature,
theoretically predicted in the case of infinite matrix
for a dimensionless radial distance $\Upsilon=0.5$ (Fig. \ref{fullfield_1}, left), it can be noted  that a strong reduction (about a factor 215 with respect to the symmetric case $\Upsilon=0$) in the modulus of $K_{\text{III, FE}}(k=2)$
is attained  for specific values of $\Upsilon$, slightly smaller than the theoretically predicted value $\Upsilon=1/2$ for the infinite plane.
Moreover, convergence of the numerical values $K_{\text{III, FE}}(k)$ to the values $K_{\text{III}}(k)$ obtained for the infinite domain is also observed at increasing the elastic matrix size.

The reliability of the SIFs at the two tips of a \lq standard' crack within an infinite matrix, eqn (\ref{sif_crack_n2}), can be also assessed in Tab. 2 through the comparison with the values reported in \cite{iran2} for circular cross sections containing multiple cracks, obtained  by numerically solving an integral equation. In particular, the numerical values of $K_{\text{III}}(k)/\left(\sqrt{\pi}\mu \Theta a^{\frac{3}{2}}\right)$ for a \lq standard' crack non-centered within a circular bar of diameter $D=20\Upsilon a/3$ are obtained starting from the values reported in Tab. 1 of \cite{iran2}. Moreover, the four values of SIFs at the two crack tips for the two cases $\Upsilon=0.75$ and $\Upsilon=1.5$ are also reported as empty circles in Fig. \ref{Knullity_crack_Circle}. For this geometry it is observed that the error provided by using the analytic expression (\ref{sif_crack_n2}) decreases at increasing values of $\Upsilon$, being this  parameter also related to both the cross section size and the distance $\left(7\Upsilon/3-1\right)a$ between the crack tip $k=1$ and the cross section boundary, which is $\left\{3/4, 4/3, 5/2, 6\right\}a$ respectively for $\Upsilon=\left\{0.75,1,1.5,3\right\}$.

\begin{table} [!ht]
\label{tab2}
\begin{center} 
\begin{tabular}{c|cc|cc}
\toprule
& \multicolumn{2}{c}{$\dfrac{K_{\text{III}}(k=1)}{\sqrt{\pi}\mu \Theta a^{\frac{3}{2}}}$} & \multicolumn{2}{c}{$\dfrac{K_{\text{III}}(k=2)}{\sqrt{\pi}\mu \Theta a^{\frac{3}{2}}}$} \\[4mm]
$\Upsilon$  & \footnotesize{infinite domain} &  \footnotesize{circular domain} & \footnotesize{infinite domain} &  \footnotesize{circular domain}  \\[.5mm]
& \footnotesize{eqn (\ref{sif_crack_n2})} & \footnotesize{$\frac{D}{2a}=\frac{10\Upsilon}{3}$ \cite{iran2}} &\footnotesize{eqn (\ref{sif_crack_n2})} & \footnotesize{$\frac{D}{2a}=\frac{10\Upsilon}{3}$ \cite{iran2}}
\\[6mm]
\hline\\
0.75            &     1.25          & 1.35769  & -0.25 & -0.324886   \\[5mm]
1          &  1.5     & 1.56797  & -0.5 & -0.554152    \\[5mm]
1.5           &   2           & 2.04046  & -1 &  -1.03536   \\[5mm]
3     &   3.5          & 3.51865  & 2.5  & -2.51746   \\
\bottomrule
\end{tabular}
\bf \small \caption{
\textnormal{
SIFs at the two tips of a \lq standard' crack evaluated from the analytical expression (\ref{sif_crack_n2}) obtained for an infinite elastic domain and the values from Tab. 1 in \cite{iran2} numerically obtained for a circular cross section of diameter $D=20\Upsilon a /3$ having a non-centered crack with $\Upsilon=\left\{0.75,1,1.5,3\right\}$. 
}}
\end{center}
\end{table}

\paragraph{A six-cusped hypocycloidal shaped void at varying the centroid position, $\Upsilon\neq0$.}

\begin{figure}[!htb]
  \begin{center}
\includegraphics[width=10 cm]{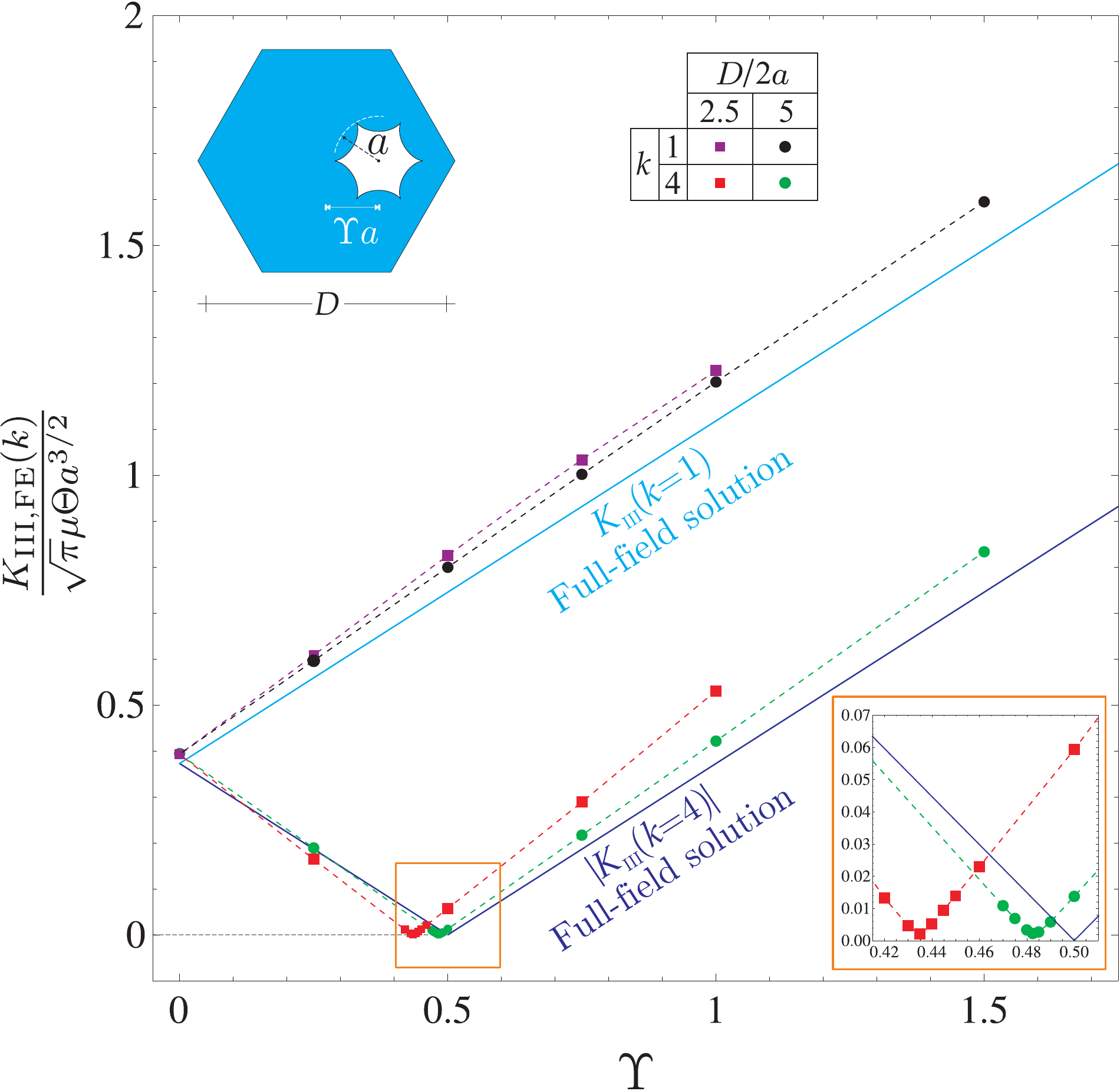}
\caption{
\small Comparison between analytical and numerical evaluation of the SIF at the first ($k=1$) and fourth ($k=4$) cusp of  a six-cusped hypocycloidal shaped hole in an elastic domain  at varying the distance parameter $\Upsilon$. Numerical evaluation is performed for an elastic matrix with
hexagonal boundary and different size ruled by the parameter $D$.
Similarly to Fig. \ref{Knullity_crack_Circle}, a strong reduction (about 160 times) in the modulus  of $K_{\text{III, FE}}(k=4)$ is observed for specific values of the distance parameter $\Upsilon \backsimeq0.5$ as the practical realization of the stress singularity
removal theoretically predicted from the presented full-field solution.
}
\label{Knullity_sixcusped_hexagon}
 \end{center}
\end{figure}

The SIFs at the first ($k=1$) and fourth ($k=4$) cusp of a six-cusped hypocycloidal shaped hole in an elastic domain are reported in Fig. \ref{Knullity_sixcusped_hexagon}
as evaluation of $K_{\text{III}}(k)$ from the analytical expression (\ref{eq_sif_hypo}) obtained for an infinite domain and of $K_{\text{III, FE}}(k)$ from the numerical simulations for an hexagonal domain of size defined by the length $D$.
Similarly to the previous example, a strong reduction about a factor 160 with respect to the symmetric case ($\Upsilon=0$) in
the modulus of  $K_{\text{III, FE}}(k=4)$ is attained  for values of $\Upsilon$ slightly smaller than 1/2, which is the  theoretically predicted value in the case of the infinite plane.

\section{Conclusions}
The full-field solution has been obtained for the torsion problem of an infinite cross section containing a void with the shape of an ellipse, an hypocycloid, or an isotoxal star-shaped polygon.
The achieved solution  has allowed for the analytical evaluation of the factors (SCF, SIF and NSIF) ruling the intensification of the shear stress in the presence of the void.
Special locations of the void have been identified for which
the stress field displays peculiar features, such as the stress annihilation at some points along the elliptical void boundary and the stress singularity removal at the cusps/points of the hypocycloidal
shaped/isotoxal star-shaped polygonal void.
Towards the application of the present model to the mechanical design of finite domain realizations,
 the reliability of the derived analytical expressions is assessed  through comparison with the numerical results obtained for specific geometries,
  showing the shape and size properties of the cross section for which the closed-form expressions provide  highly accurate predictions.

\section*{Acknowledgments}
The authors gratefully acknowledge financial support from the ERC Advanced Grant \lq Instabilities and nonlocal multiscale modelling of materials'
ERC-2013-ADG-340561-INSTABILITIES.

\newpage

\setcounter{equation}{0}
\renewcommand{\theequation}{{A}.\arabic{equation}}
\begin{center}
{\bf APPENDIX A - Evaluation of Stress Intensification Factors}\label{appconvergence}\\
\end{center}

\begin{figure}[h!]
  \begin{center}
\includegraphics[width=10 cm]{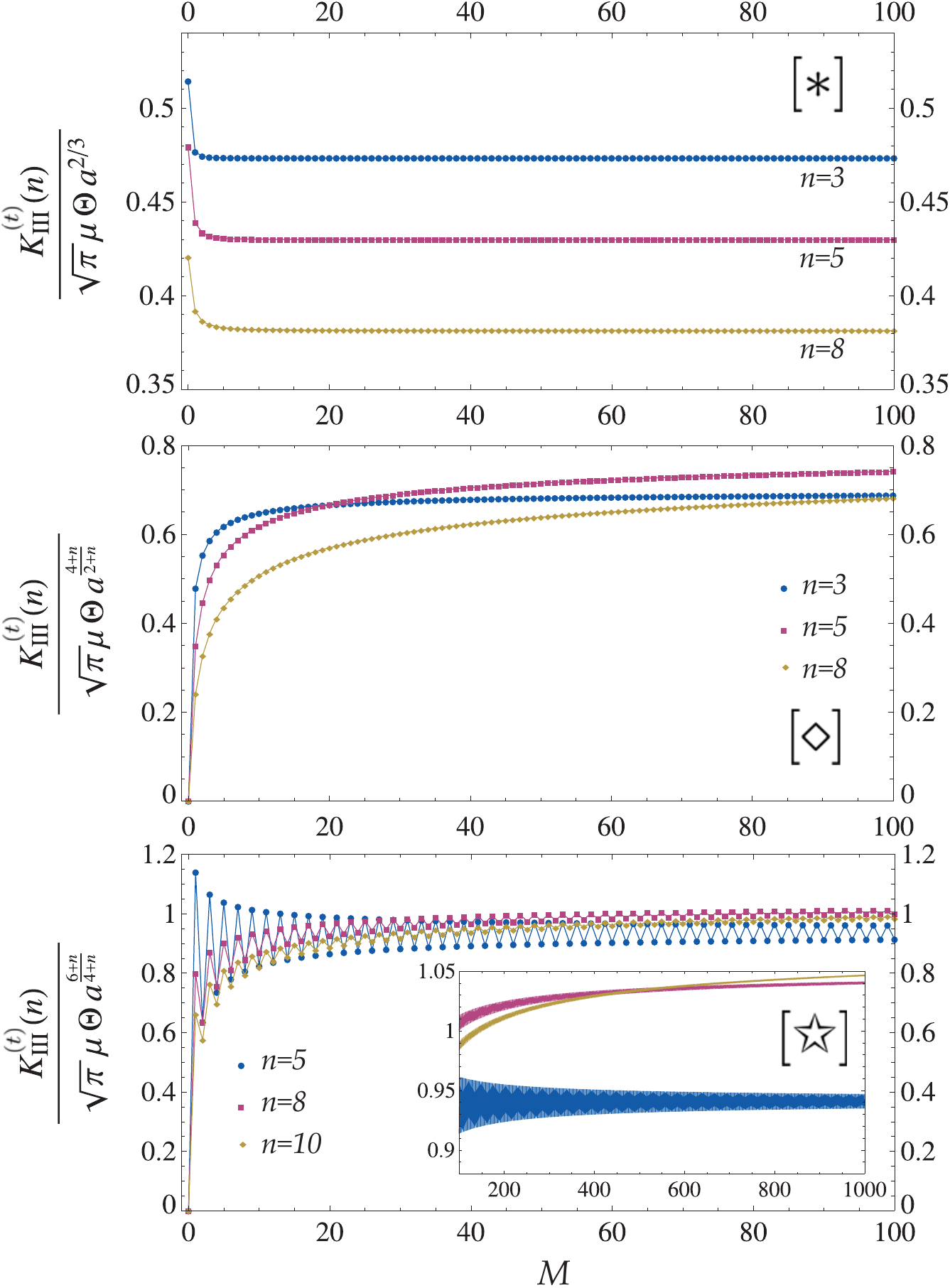}
\caption{ Torsional contribution of the Stress Intensity Factor, $K^{(t)}_{\text{III}}$, for $n$-pointed 
star-shaped cracks (upper part), for $n$-sided regular polygonal voids, and for $n$-pointed regular star polygonal voids as the  approximation of the respective series by the finite sum of its first $M$ terms.
The convergence rate is displayed at different values of the point number $n$ for the three void shapes, each of these showing a different order ($M\simeq 10$ for $n$-pointed star-shaped cracks, $M\simeq 100$ for $n$-sided regular polygonal voids, and $M\simeq 1000$ for $n$-pointed regular star polygonal voids).
\small
}
\label{convergenza}
 \end{center}
\end{figure}

Differently from the uniform Mode III  contribution $K_{III}^{(u)}$, the torsional contribution $K_{III}^{(t)}$ in SIFs and NSIFs involves the evaluation of $\mathcal{C}^{\left[\cdot\right]}(n)$, eqn (\ref{matcalCC}), which requires the computation of a series
in the case of isotoxal star-shaped polygonal voids.
The convergence in the evaluation of $K_{III}^{(t)}$ through the truncation of the series and its approximation by the finite sum of its first $M$ terms is shown Fig. \ref{convergenza}.
It can be observed that the number $M$ of terms of the finite sum needed to well approximate the series dramatically changes at varying the void geometry. In particular, a satisfactory convergence  is reached for $M \simeq 10$ for star-shaped cracks (Fig. \ref{convergenza}, upper part),
for $M \simeq 100$ for polygonal voids (Fig. \ref{convergenza}, central part) and for $M \simeq 1000$ for star-shaped polygonal voids (Fig. \ref{convergenza}, lower part), for which a peculiar oscillatory behaviour is also observed.
\end{document}